\documentclass[aps,prd,twocolumn,groupedaddress,showpacs,nofootinbib]{revtex4}
\usepackage{graphicx,dcolumn,bm,amssymb,amsmath,latexsym,amsfonts,footnote}
\usepackage[usenames]{color} 
\usepackage{float}
\begin{document}

\newcommand{\nonu}{\nonumber}

\newcommand{\sm}{\small}
\newcommand{\noi}{\noindent}
\newcommand{\npg}{\newpage}
\newcommand{\nl}{\newline}
\newcommand{\bp}{\begin{picture}}
\newcommand{\ep}{\end{picture}}
\newcommand{\bc}{\begin{center}}
\newcommand{\ec}{\end{center}}
\newcommand{\be}{\begin{equation}}
\newcommand{\ee}{\end{equation}}
\newcommand{\beal}{\begin{align}}
\newcommand{\eeal}{\end{align}}
\newcommand{\bea}{\begin{eqnarray}}
\newcommand{\eea}{\end{eqnarray}}
\newcommand{\bnabla}{\mbox{\boldmath $\nabla$}}
\newcommand{\univec}{\textbf{a}}
\newcommand{\VectorA}{\textbf{A}}
\newcommand{\Pint}

\title{Remarks on electrical Penrose process for magnetized Reissner-Nordstr\"om black hole}

\author{A. Baez$^{1,}$\footnote{jose.baez@cinvestav.mx}, Nora Breton$^{1,}$\footnote{nora.breton@cinvestav.mx}, and I. Cabrera-Munguia $^{2,}$\footnote{icabreramunguia@gmail.com}}
\affiliation{$^{1}$ Departamento de F\'isica, Centro de Investigaci\'on y de Estudios
 Avanzados del Instituto Politecnico Nacional; Apdo. Postal 14-740, Mexico City, Mexico\\
$^{2}$Departamento de F\'isica y Matem\'aticas, \\
Universidad Aut\'onoma de Ciudad Ju\'arez, 32310 Ciudad Ju\'arez, Chihuahua, M\'exico}



\begin{abstract}

The energy extraction from a magnetized Reissner-Nordstr\"om black hole is analyzed within the framework of the electric Penrose mechanism. The presence of an external magnetic field induces an axisymmetric configuration and an ergosphere (the region where energy extraction is possible) arises, allowing for negative energy states even in an otherwise static spacetime. By analyzing the decay of particles at turning points of the radial motion, we derive the general expression for the efficiency of the process in terms of the metric coefficients and the electromagnetic potential. The resulting efficiencies are interpreted as Killing energy efficiencies associated with the local splitting process, while the escape of the positive energy fragment is treated as an additional effective-potential requirement. This formulation provides a direct criterion for identifying the ergoregions  and we show that the magnetic field acts as a control parameter that governs both the configuration of the ergosphere and the efficiency of the process. In particular, analytical expressions for the critical magnetic fields that determine the onset and suppression of energy extraction are determined. Our results extend previous analysis of the electric Penrose process for magnetized configurations and clarify the role of the external field in enhancing or inhibiting energy extraction from charged black holes.
\end{abstract}

\maketitle

\section{Introduction}
The extraction of energy from black holes (BHs),  has been the subject of continuous interest since the seminal proposal of the Penrose process \cite{PenFloyd1971}, originally conceived for rotating BHs like the Kerr spacetime \cite{Kerr1963,Visser2007}. In its classical formulation {for an asymptotically flat spacetime}, an ongoing particle undergoes a decay process, producing a fragment with negative energy (as measured at infinity) that is captured by the BH and then, due to energy conservation, the second fragment escapes to infinity with enhanced energy. This mechanism relies fundamentally on the existence of an ergoregion, a feature absent in static geometries.

Nevertheless, the energy extraction process is not restricted to rotating BHs, the electric Penrose process provides an alternative mechanism in which energy can be extracted from static, charged BHs through electrostatic interactions between the background electromagnetic field and charged test particles of opposite charge with respect to the BH charge \cite{Christo1971,DenardoRuffini1972,Dadich1980}. In this case, the negative energy trajectories necessary for the energy extraction arise due to the coupling of the test charge with the electromagnetic potential rather than due to frame dragging. The general properties of this process, including the conditions for negative energy states and energy extraction, have been analyzed in different settings, for example \cite{Zaslavski2024}. More recent studies include extensions to asymptotically de Sitter and anti-de Sitter spacetimes \cite{Lemos2024, Baez2024,Sourav2018} and to multiple configurations of BHs such as Majumdar-Papapetrou and Bonnor spacetimes \cite{Richartz2021,Baez2022}.

In astrophysical scenarios, BHs are often immersed in electromagnetic environments where magnetic fields play a crucial role \cite{Kolos2015}. The presence of an external  magnetic field can significantly modify the dynamics of charged particles near the BH \cite{Kolos2021}. In particular, magnetized BH solutions exhibit qualitatively new features, such as the emergence of regions where energy extraction is possible (ergoregions) even in static configurations  \cite{Gibbons2013}. This is the case of the magnetized Reissner-Nordstr\"om spacetime, where the external magnetic field induces an axisymmetric structure, generating a generalized ergosphere. Therefore, as a consequence, energy extraction mechanisms may arise from the interplay between gravitational and electromagnetic effects \cite{Nucamendi2022}.

Moreover, high-energy astrophysical phenomena can admit a possible explanation based on variations of the Penrose process. For instance, variants of the Penrose process have been proposed as mechanisms for particle acceleration to arbitrarily high center-of-mass energies \cite{Schnittman2018,BSW2009}, as well as for the generation of high-energy emissions. Related developments involving rotational energy extraction, superradiance and locally non rotating frames can be found in \cite{Bardeen1972, Wald1974, Starobinsky1973,Teu1974,Damour75}. Moreover, the presence of external magnetic fields surrounding  rotating BHs give
rise to accretion disks comprising charged ionized matter \cite{Kolos2021}; this could potentially be connected to the observation of high-frequency oscillations observed in microquasars or galactic nuclei where ultra high-energy particles around rotating magnetized BHs are created \cite{Kolos2017, Kolos2015}. Additionally, in magnetized environments, radiative processes associated with charged particle motion may lead to observable signatures such as synchrotron radiation and quasi-periodic oscillations \cite{Kolos2017,Kolos2015,Kolos2018, KolosPRD2021}. In particular, electromagnetic versions of the Penrose process have been shown to enhance efficiency beyond the purely rotational limit \cite{Bhat1985,Tursunov2021,Wagh1985,Wagh1989,Kolos2020,Nucamendi2022}. 
It is worth mentioning that the mechanisms considered here are entirely classical and differ fundamentally from quantum processes of BH energy release such as Hawking radiation \cite{Damour75}, since they do not require a time-dependent background geometry.

The present paper aims to investigate the energy extraction process in the context of a magnetized Reissner-Nordstr\"om BH, extending and elucidating some aspects presented in \cite{Nucamendi2022}. Our approach is based on the analysis of uncharged/charged particle decay at turning points of the radial motion, which allows us to obtain explicit expressions for the energies of the fragments  and for the efficiency in terms of the background geometry and the electromagnetic potential.  This formulation makes explicit the interplay between gravitational and electromagnetic contributions for the energy balance and provides  natural criteria to identify the regions where negative energy trajectories (and therefore energy extraction) are possible.

Special attention is devoted to the role of the external magnetic field as a control parameter of the process. In particular, we provide an analytic characterization of the critical magnetic fields that determine the existence and structure of the ergoregion, as well as the onset and suppression of energy extraction. The ergoregion is delimited by the horizon and the static limit surface ($SLS$) and our analysis reveals a close relationship between the behavior of the $SLS$  and the efficiency of the process, allowing us to identify configurations in which the extraction region is maximized or completely suppressed.

A common simplifying assumption in many studies of energy extraction processes is to focus on the behavior of particles at the event horizon. While this approach provides valuable insight into the maximum efficiency of the process, it does not fully capture the structure of the region where negative energy states can occur. In particular, the
boundary of the extraction region typically associated with the surface of limiting static observers plays a crucial role in determining where energy extraction is actually possible.

In this work, we go beyond the horizon-based analysis and perform a systematic study of the extraction region in the magnetized Reissner-Nordstr\"om spacetime. We show that the location and extent of this region are not fixed by the background geometry alone, but can be significantly modified by the parameters of the particles involved
in the process. In particular, the charges of the fragments introduce additional degrees of freedom that can either enlarge or reduce the region where negative energy states are allowed, leading to a richer structure of the parameter space.

From a broader perspective, this result suggests a natural connection between energy extraction mechanisms and dynamical processes in strong gravitational and electromagnetic fields. Since the existence of negative-energy states is closely tied to the effective radial potential governing particle motion, the analysis presented here opens the possibility of linking extraction processes with orbital dynamics and particle trajectories in magnetized spacetimes.

Furthermore, these considerations suggest a possible connections between different energy extraction approaches. In particular, the role of electromagnetic fields in shaping the extraction region suggests a conceptual connection with mechanisms such as the Blandford-Znajek process \cite{BZ1977} and magnetic reconnection scenarios as discussed by Comisso and Asenjo \cite{Comisso2021}. While the underlying physical setups differ, the common ingredient is the interplay between electromagnetic fields and spacetime geometry in enabling the extraction of energy from compact objects.

This paper is organized as follows. In Section \ref{sec2}, we introduce the general formalism for charged particle motion in stationary axially symmetric charged BH and derive the conditions for the existence of energy extraction region. In Section \ref{MRN}, we specialize the analysis to the magnetized Reissner-Nordstr\"om spacetime and discuss the role of the external magnetic field in determining the structure of the extraction region and static limit surface. In Section \ref{secEffi}, we analyze the efficiency of the {local} extraction process and its parameter space structure for neutral and charged particles, including the characterization of critical magnetic fields and charges configurations {also we discuss the kinematic conditions for the escaping particle}. Finally, in Section \ref{sec4}, we summarize our results and discuss possible extensions.

\section{Electric Penrose mechanism for stationary axisymmetric charged black holes}
\label{sec2}
\vspace{-0.3cm}
This Section is devoted to elucidate the energy extraction from a stationary axisymmetric charged BH.
Following  the electric Penrose process as introduced in \cite{PenFloyd1971}, we approach the study of its efficiency in the vicinity of an axisymmetric charged BH. {The electric Penrose mechanism consists in sending a charged particle towards the BH; at some point, inside the generalized ergosphere, the particle breaks up into two fragments, one of them remains inside the ergosphere and eventually penetrates the horizon of the BH, while the second one propagates outward from the extraction region, or to infinity in asymptotically flat cases, carrying more conserved energy than the initial one.}

Let us consider the stationary axisymmetric line element,
\begin{equation}\label{stametric}
ds^2 = g_{tt} dt^2 +2 g_{t\phi} dt d\phi+g_{rr} dr^2 + g_{\theta\theta}d\theta^2 +g_{\phi\phi}d\phi^2,
\end{equation}
\noi where $(t, r,\theta,\phi)$ are the Boyer-Lindquist coordinates used to describe the Kerr BH; the metric components $g_{\mu\nu}$  depend only on the coordinates $r$ and $\theta$. We use the signature $(-,+,+,+)$. 
The electromagnetic potential $A_{\mu}$ of the charged BH and the magnetic external field has two nonvanishing components,

\begin{equation}
A_{\mu}=(A_t, 0,0,A_\phi).
\end{equation}

The Penrose process is strongly dependent  on the trajectories of the test particles that penetrate the ergosphere. Such trajectories are analyzed in the next subsection.
\subsection{Motion of charged particles}\label{sec21}

The  equation of motion for a test particle with charge $q$,  mass $m$, and charge-mass ratio  $e=q/m$,  can be obtained from the Euler-Lagrange equations with the Lagrangian 
\begin{equation}
\mathcal{L} = \frac{1}{2} g_{\mu\nu}\dot{x}^{\mu}\dot{x}^{\nu} + e A_{\alpha}\dot{x}^{\alpha},
\end{equation}
\noi where the dot means derivative with respect to an affine parameter $\lambda$. For a massive particle, the affine parameter is $\lambda=\tau/m$, where $\tau$ is the proper time.  In the case of a massless particle $\lambda$ is a properly chosen affine parameter. In  terms of the metric functions in Eq. (\ref{stametric}), the Lagrangian is

\begin{equation}
\begin{aligned}
\mathcal{L} = & \frac{1}{2}\left(g_{tt}\dot{t}^2+2g_{t\phi} \dot{t}\dot{\phi}+g_{rr}\dot{r}^2+g_{\theta\theta}\dot{\theta}^2+g_{\phi\phi}\dot{\phi}^2\right)\\ &\qquad +e A_{t}\dot{t}+e A_{\phi}\dot{\phi}.
\end{aligned}
\end{equation}

\noi Since $\mathcal{L}$  does not depend explicitly on $(t,\phi)$ then we can identify two motion constants of the test particle: its energy and its angular momentum per unit mass,   $\mathcal{E}$ and $l$,  respectively, given by
\begin{equation}
\begin{aligned}
l & =\frac{L}{m}=\frac{\partial\mathcal{L}}{\partial\dot{\phi}}=g_{\phi\phi}\dot{\phi}+g_{t\phi} \dot{t}+e A_\phi, \\
\mathcal{E} & =\frac{E}{m}=-\frac{\partial\mathcal{L}}{\partial\dot{t}}=-g_{tt}\dot{t}-g_{t\phi} \dot{\phi}-eA_{t}.
\end{aligned}
\end{equation}
\noi where $E$ and $L$ are the energy and angular momentum, respectively,
of the test particle. The  motion equations are obtained by solving for $\dot{\phi}$ and $\dot{t}$, 
\begin{equation}
\begin{aligned}
\dot{\phi} & =\frac{-g_{t\phi}\left(\mathcal{E}+e A_t\right)-g_{tt}\left(l-e A_\phi\right)}{g_{t\phi}^2-g_{tt}g_{\phi\phi}},\\  \dot{t} & =\frac{g_{\phi\phi}\left(\mathcal{E} +eA_{t}\right)+g_{t\phi}\left(l-e A_{\phi}\right)}{g_{t\phi}^2-g_{tt}g_{\phi\phi}}.   
\end{aligned}
\end{equation}

\noi Since the metric (\ref{stametric}) is invariant under transformation $\theta = \pi -\theta$, that is, there is mirror symmetry with respect to the equatorial plane, we can confine the movement of the particles to $\theta=\pi/2$ and $\dot{\theta}=0$. Then, using the contraction of the four-momentum $\dot{x}^\mu \dot{x}_{\mu}=-\delta$, we obtain the $r$- component of the velocity, $\dot{r}$, as

\begin{equation}\label{geodesics}
\begin{aligned}
\dot{r}^2 =\frac{1}{g_{rr}} & \Bigg[\frac{g_{\phi\phi}}{g_{t\phi}^2-g_{tt}g_{\phi\phi}}\left(\mathcal{E}+e A_{t}+\frac{g_{t\phi}\left(l-e A_{\phi}\right)}{g_{\phi\phi}}\right)^2\\ & -\frac{\left(l-e A_{\phi}\right)^2}{g_{\phi\phi}}-\delta\Bigg],
\end{aligned}
\end{equation}
where $\delta=1$ for massive particles and $\delta=0$ for massless particles.

\noi For the $i$-particle the turning points of the radial motion, $\dot{r}=0$, are determined by the equation

\begin{equation}\label{conditioneg}
\begin{aligned}
&\quad\big[g_{\phi\phi}\left(\mathcal{E}_i+e_i A_t\right) + g_{t\phi}\left(l_i-e_i A_\phi\right)\big]^2\\  &{-\left(g_{t\phi}^2-g_{tt}g_{\phi\phi}\right)}\left[\left(l_i-e_i A_\phi\right)^2+\delta_i g_{\phi\phi}\right]=0,
\end{aligned}
\end{equation}

\noi that solving for $\mathcal{E}_i$ give
\begin{equation}\label{potefectivo}
\begin{aligned}
&\qquad\mathcal{E}_{i}=V_{\pm i}=-e_i A_t-\left(l_i-e_i A_\phi\right)\frac{g_{t\phi}}{g_{\phi\phi}}\\
&\pm\frac{\sqrt{\left(\left(l_i-e_i A_\phi\right)^2+\delta_i g_{\phi\phi}\right)\left(g_{t\phi}^2-g_{tt} g_{\phi\phi}\right)}}{g_{\phi\phi}},
\end{aligned}
\end{equation}
where $V_{\pm i}$ is the effective potential for the $i$- charged test particle; we denote the test particles by $i=0,  1, 2$  the initial particle, the one that penetrates the BH and the one that escapes from the BH, respectively. In what follows, we analyze the potential 
$V_{\pm  1}$, corresponding to particle 1, that penetrates the BH horizon.


\subsection{Energetic conditions for the extraction process}\label{pp}

Let us consider a charged test particle moving along a timelike geodesic with energy and angular momentum ($E_0>0$, $L_0$), that reaches one turning point ($\dot{r}=0$); at that point it breaks up into two pieces, with energies and angular momentum ($E_1$, $L_1$) and ($E_2$, $L_2$), such that $E_1<0$ and $E_2>0$. The particle with negative energy (particle 1) is confined within the generalized ergosphere until it falls into the BH, while the other particle (particle 2) escapes from the generalized ergosphere with more energy than the initial particle, $E_2>E_0$. 

The trajectory of the incident particle (particle 0), $x_0^\mu (\lambda)$, is timelike, parametrized by $\lambda=\tau/m$ where $\tau$ is the proper time; this trajectory initially is outside the ergosphere and ends inside it at the break-up point ($r_*$,$\theta_*$,$\phi_*$). From the break-up point emerge two particles (1 and 2), which can move in timelike or lightlike trajectories, depending on the type of splitting; these trajectories $x_{1,2}^\mu (\lambda)$ are parametrized by the proper time or affine parameter, respectively. We denote the mass, charge mass ratio, energy per unit mass,  azimuthal angular momentum per unit mass, and  4-momentum of the $i$- particle by $m_i$, $e_i$, $\mathcal{E}_i$, $l_i$ and $p_i^\mu = dx^\mu_i/d\lambda$, respectively. These quantities must fulfill the charge and 4-momentum conservation, 
\begin{equation}\label{qconservation}
m_0 e_0=m_1 e_1+m_2 e_2,
\end{equation}
and
\begin{equation}\label{4momentum}
 p_0^\mu=p_1^\mu+p_2^\mu.
\end{equation}

Moreover, at the break-up point, the conserved canonical energy associated with stationarity is preserved, while the azimuthal component gives the conservation of angular momentum, namely,
\begin{align}
m_0 \mathcal{E}_0=m_1 \mathcal{E}_1+m_2 \mathcal{E}_2,\label{Econservation}\\
 m_0 l_0=m_1 l_1+m_2 l_2.\label{Lconservation}  
\end{align}

{To determine the efficiency of the Penrose process, we need to know the explicit form of the energies $E_i$ of each particle at the break-up point, which we have considered as a turning point. The results derived below apply to a restricted but analytically tractable class of splitting events, those occurring at common radial turning points of the parent and daughter particles, namely
\begin{equation}
    \dot{r}_0=\dot{r}_1=\dot{r}_2=0.
\end{equation}
Therefore, the energetics conditions and efficiency obtained should be interpreted for this local turning point kinematics and not as the most general Penrose process. From Eq. \eqref{conditioneg}, we obtain the angular momentum for the particle $i$ at the break-up point, given by
\begin{equation}\label{angularmomentumi}
\begin{aligned}
    l_i=&e_i A_\phi-\left(\mathcal{E}_i+e_i A_t\right)\frac{g_{t\phi}}{g_{tt}}\\
        &+\epsilon_i \frac{\sqrt{\left(\left(\mathcal{E}_i+e_i A_t\right)^2+\delta_i g_{tt}\right)\left(g_{t\phi}^2-g_{tt} g_{\phi\phi}\right)}}{g_{tt}},
\end{aligned}
\end{equation}
where $\epsilon_i=\pm1$. Using Eqs. \eqref{Lconservation} and \eqref{angularmomentumi} we obtain,
\begin{equation}\label{energysystem}
    \begin{aligned}
        &\epsilon_0 m_0 \sqrt{\left(\mathcal{E}_0+e_0 A_t\right)^2+\delta_0 g_{tt}}\\
        +&\epsilon_1 m_1 \sqrt{\left(\mathcal{E}_1+e_1 A_t\right)^2+\delta_1 g_{tt}}\\
        +&\epsilon_2 m_2 \sqrt{\left(\mathcal{E}_2+e_2 A_t\right)^2+\delta_2 g_{tt}}=0. 
    \end{aligned}
\end{equation}
It is worth mentioning that the product $e_i A_{\phi}$ does not appear explicitly in Eq. \eqref{energysystem}, independently of the choice of the signs $\epsilon_i$. Thus, the different sign choices only distinguish the corresponding radial branches and do not modify the explicit dependence of the energy condition on the electromagnetic potential. The horizon values, when considered, are understood as limiting values.}

Solving the system of Eqs.\ \eqref{Econservation} and \eqref{energysystem} we obtain $\tilde{E}_1$ and $\tilde{E}_2$ in terms of the  energy of the incident particle, $\tilde{E}_0$

\begin{equation}\label{energies}
\begin{aligned}
&\tilde{E}_1=\frac{1}{2} \left(\left(1+\frac{\tilde{\delta}_1}{\tilde{\delta}_0}-\frac{\tilde{\delta}_2}{\tilde{\delta}_0}\right)\tilde{E}_0+\kappa\sqrt{d_0}\right),\\
&\tilde{E}_2=\frac{1}{2}\left(\left(1-\frac{\tilde{\delta}_1}{\tilde{\delta}_0}+\frac{\tilde{\delta}_2}{\tilde{\delta}_0}\right)\tilde{E}_0-\kappa\sqrt{d_0}\right),\\
& \tilde{E}_i=E_i+q_i A_t, \quad i=1,2, \quad \kappa=\pm 1,\\
d_0=& \left( 1- 2 \left( \frac{\tilde{\delta}_1+\tilde{\delta}_2}{\tilde{\delta}_0}\right) + \left( \frac{\tilde{\delta}_1-\tilde{\delta}_2}{\tilde{\delta}_0} \right)^2 \right)
\left(\tilde{E}_0^2+g_{tt}\tilde{\delta}_0 \right),
\end{aligned}
\end{equation}

\noi where $E_i=m_i \mathcal{E}$, $q_i=m_i e_i$ are the energy and charge of the $i$-particle, and $\tilde{\delta}_i=m^2_i \delta_i$, $\tilde{\delta}_i$ is $m^2_i$ for massive particles and zero for massless particles. Note that Eq.\ (\ref{energies}) does not depend on the sign of $l_i$ in Eq. (\ref{conditioneg}); nevertheless, the sign must be considered to determine the viability of the energy extraction process. Additionally, note that Eq.\ (\ref{energies}) is of the same form as the solutions presented in \cite{Baez2024} for a static and spherically symmetric BH, that is, Eq.\ (\ref{energies}) does not depend on $g_{t \phi}$ .

\section{Magnetized Reissner-Nordstr\"om BH}\label{MRN}

For completeness  in this section we review some results on the Penrose process for the magnetized Reissner-Nordstr\"om (RN) BH. This spacetime describes an axisymmetric magnetized RN BH, that in coordinates ($t$, $r$, $\theta$, $\phi$) is given by,

\begin{equation}
\label{rnmmetric1}
\begin{aligned}
ds^2 =&H (-F dt^2+F^{-1} dr^2+r^2 d\theta^2)+H^{-1} r^2\sin^2\theta\\
&\times \left(d\phi-\omega dt\right)^2,
\end{aligned}
\end{equation}
where
\begin{equation}\label{rnmmetric2}
\begin{aligned}
F&=1-\frac{2M}{r}+\frac{Q^2}{r^2},\\
H&=1+\frac{1}{2}B^2\left(r^2\sin^2\theta+3Q^2\cos^2\theta\right)\\
&+\frac{1}{16}B^4\left(r^2\sin^2\theta+Q^2\cos^2\theta\right)^2,\\
\omega &=-\frac{2Q B}{r}+\frac{1}{2}QB^3 r\left(1+F\cos^2\theta\right),
\end{aligned}
\end{equation}
characterized by the BH mass and charge, $M$ and $Q$, and the uniform external magnetic field $B$. The event horizon is given by $F=0$, 
\begin{equation}
r_{+}=M+\sqrt{M^2-Q^2},
\end{equation}
note that it does not depend on the external field $B$, and the horizon values lie in $M \leq r_{+} \leq 2M$, corresponding to the extreme RN case, $M=Q$, and the Schwarzschild BH ($Q=0$), respectively.\\
The nonvanishing components of the electromagnetic potential are
\begin{equation}
    A = \Phi_0 dt +\Phi_3 \left(d\phi-\omega dt\right),
\end{equation}
with \cite{Gibbons2013,Shayma2021},
\begin{equation}\label{rnmpot}
    \begin{aligned}
        \Phi_0&=-\frac{Q}{r}+\frac{3}{4} Q B^2 r (1+F \cos^2\theta),\\
        \Phi_3 & =\frac{2}{B}-H^{-1}\left(\frac{2}{B}+\frac{1}{2}B\left(r^2\sin^2\theta+3Q^2\cos^2\theta\right)\right).
    \end{aligned}
\end{equation}

To extend the analysis  presented in \cite{Gibbons2013, Nucamendi2022} we discuss some properties of the ergoregion in the equatorial plane. 
{The ergoregion is the region where energy can be extracted and it is bounded or delimited by the BH horizon and the static limit surface ($SLS$)}. The $SLS$ is determined by the condition $g_{tt}=0$, which  from Eqs.\ (\ref{rnmmetric1})- (\ref{rnmmetric2}),
\begin{equation}\label{gtt}
    g_{tt}=-H F +H^{-1}r^2\omega^2 \sin^2\theta=0.
\end{equation}
In the equatorial plane, $\theta=\pi/2$, this yields the condition
\begin{equation}\label{gttcondition}
    -H^2 F +r^2 \omega^2=0,
\end{equation}
or, explicitly,
\begin{equation}\label{gttcondition2}
    (2M-r)r\left(B^2 r^2+4\right)^4-Q^2\left(B^4 r^4-24B^2 r^2+16\right)^2=0.
\end{equation}
Eq.\ (\ref{gttcondition2}) is complicated to solve analytically for $r$; nevertheless, we can characterize the $SLS$ in terms of the magnetic field $B$ for a fixed radius and delimit the extraction region. Therefore, from Eq.\ (\ref{gttcondition2}) after some straightforward algebra, we found the physically acceptable solutions for the magnetic field  as,
\begin{equation}\label{Bcondition0}
    B\left(r,a, Q, \epsilon \right) = \frac{2}{r}\frac{\sqrt{Q-a}}{\sqrt{2Q}+\epsilon \sqrt{Q+a}}, \qquad \epsilon=\pm 1,
\end{equation}
where $a=\pm\sqrt{\left(2M-r\right)r}$; this restricts the range of the radial coordinate for the $SLS$ to $r_+\leq r_{eg}\leq 2M$ and the values of $a$ to the domain $-Q \leq a < Q$. Depending on the choice of $\epsilon$, one can easily prove that the following inequalities are satisfied,
\begin{equation}\label{Bepsilons}
\begin{aligned}
   & 0\leq B\left(r,a, Q, \epsilon=1 \right)\leq \frac{2}{r_{\rm eg}},\\
   &\frac{2}{r_{\rm eg}} \leq B\left(r,a, Q, \epsilon=-1 \right)<\infty,
    \end{aligned}
\end{equation}
where $r_{\rm eg}$ is the boundary of the ergoregion.
The explicit form of solutions in Eq.\ (\ref{Bcondition0}) and the compact notation that we will use in this work are given in Table\ \ref{Tab1}.\\ 
\begin{table}[htb]
\centering
\caption{The magnetic field $B\left(r,a, Q, \epsilon \right)$, from Eq.\ (\ref{Bcondition0}), corresponding to the ergosphere boundary defined by $g_{tt}=0$. The set of parameters $a$ and $\epsilon$ are tabulated.} \label{Tab1}
\begin{tabular}{c c c c c c}
\hline
            & $\epsilon$      & $a$  &\\ \hline
$B_1\left(r\right)$    & $1$     & $\sqrt{\left(2M-r\right)r}$&\\
$B_2\left(r\right)$    & $1$       & $-\sqrt{\left(2M-r\right)r}$ &\\
$B_3\left(r\right)$    & $-1$       & $-\sqrt{\left(2M-r\right)r}$ &\\
$B_4\left(r\right)$    &   $-1$     &   $\sqrt{\left(2M-r\right)r}$     &\\
\hline
\end{tabular}
\end{table}

Besides, the field strengths $B$ in Table\ \ref{Tab1} are ordered as,
\begin{equation}
    B_1 \left(r\right)< B_2 \left(r\right)<B_3 \left(r\right)< B_4\left(r\right),
\end{equation}
this determines the sign of $g_{tt}$ according to the ranges of the magnetic field $B$, as
\begin{equation}
\begin{aligned}\label{Bcondition2}
    g_{tt}<0\quad &\mathrm{if} \quad 0<B<B_1\left(r\right),\\
    g_{tt}>0\quad &\mathrm{if} \quad B_1\left(r\right)<B<B_2\left(r\right)\\
    g_{tt}<0\quad &\mathrm{if} \quad B_2\left(r\right)<B<B_3\left(r\right),\\
    g_{tt}>0\quad &\mathrm{if} \quad B_3\left(r\right)<B<B_4\left(r\right),\\
    g_{tt}<0\quad &\mathrm{if} \quad B_4\left(r\right)<B.
\end{aligned}
\end{equation}
This is quite important because the classical Penrose process occurs inside the ergosphere, which is precisely the region where $g_{tt}>0$.

On the other hand, we can find the maximum and minimum values of the $SLS$ as well as the magnitudes of the magnetic field that generate these scenarios. Using Eq.\ (\ref{gttcondition}), deriving implicitly with respect to $B$ and taking $dr/dB =0$, we obtain the critical points (maxima and minima) and the corresponding radius of the $SLS$, the minimum and maximum radii of $SLS$ are $r_{SLS}^\mathrm{min}=r_+$ and $r_{SLS}^\mathrm{max}=2M$, respectively. Evaluating at these $r$ for the field strengths in Table\ \ref{Tab1} we obtain the magnetic fields  shown in the Tables \ref{Tab2}-\ref{Tab3}.

\begin{table}[htb]
\centering
\caption{Values of magnetic field $B\left(r,a, Q, \epsilon \right)$ at the minimum radius of $SLS$, $r_{SLS}^\mathrm{min}=r_+$.} \label{Tab2}
\begin{tabular}{c c c c c c}
\hline
             & $\epsilon$      & $a$  & $B\left(r,a, Q, \epsilon \right)$&\\ \hline
$B_1\left(r_+\right)$    & $1$     & $Q$& $0$ &\\
$B_2\left(r_+\right)$    & $1$       & $-Q$ & $2/r_+$&\\
$B_3\left(r_+\right)$    & $-1$       & $-Q$ & $2/r_+$&\\
$B_4\left(r_+\right)$    & $-1$       & $Q$ & $\infty$ &\\
\hline
\end{tabular}
\end{table}

\begin{figure}[htb]
\includegraphics[width=7cm,height=7cm]{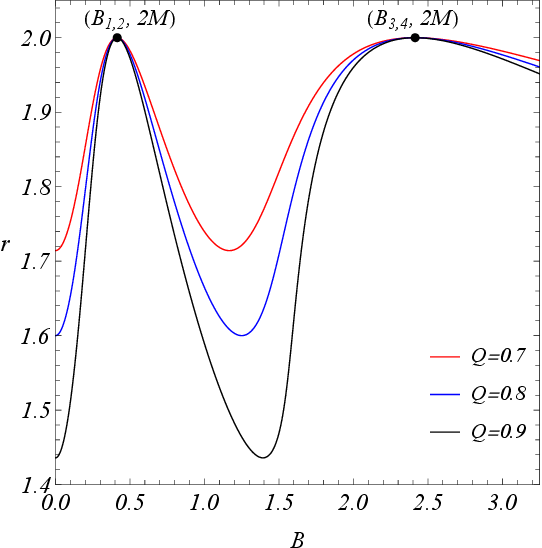}
\caption{\label{FIG1}The limit radii of  $SLS$ as a function of the magnetic field $B$ in the equatorial plane, for different values of BH charge, $Q$. The BH mass has been fixed as $M=1$. The maxima $SLS$ occurs for $r_{SLS}^{\rm max}=2M$, with magnetic field $B=B_{1,2}\left(2M\right)$ and $B=B_{3,4}\left(2M\right)$, that are indicated by the black dots.}
\end{figure}

\begin{figure}[htb]
\includegraphics[width=7cm,height=7cm]{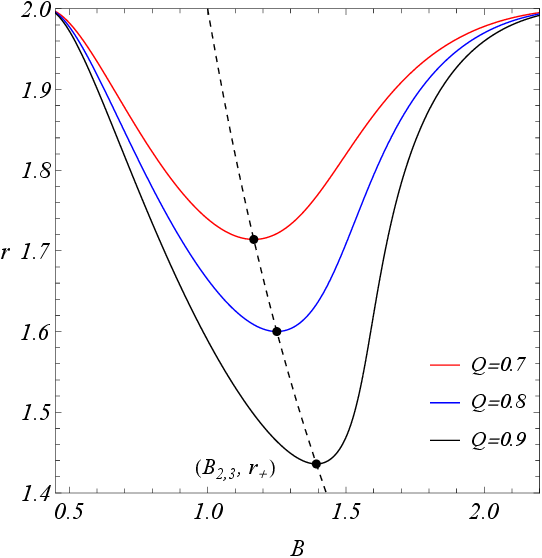}
\caption{\label{FIG2}Dependence of $SLS$ with respect to the magnetic field $B$ in the equatorial plane, for different BH charges, $Q$. The BH mass is fixed to $M=1$. The minima $SLS$ occur for $r_{SLS}^{\rm min}=r_+$ and $B=B_{2,3}\left(r_+\right)$, represented by the dashed line for each value of $Q$.}
\end{figure}

\begin{table}[htb]
\centering
\caption{Values of magnetic field $B\left(r,a, Q, \epsilon \right)$ in the maximum radius of $SLS$, $r_{SLS}^\mathrm{max}=2M$.} \label{Tab3}
\begin{tabular}{c c c c c c}
\hline
             & $\epsilon$      & $a$  & $B\left(r,a, Q, \epsilon \right)$&\\ \hline
$B_1\left(2M\right)$    & $1$     & $0$& $\left(\sqrt{2}-1\right)/M$ &\\
$B_2\left(2M\right)$    & $1$       & $0$ & $\left(\sqrt{2}-1\right)/M$&\\
$B_3\left(2M\right)$    & $-1$       & $0$ & $\left(\sqrt{2}+1\right)/M$&\\
$B_4\left(2M\right)$    & $-1$       & $0$ & $\left(\sqrt{2}+1\right)/M$ &\\
\hline
\end{tabular}
\end{table}
An aspect to highlight with respect to the magnetic field in Tables \ \ref{Tab2} and \ref{Tab3} is the degeneracy that occurs in the solutions at the event horizon and on the $SLS$.
In addition, the existence of several critical points suggests that the behavior of $SLS$ as a function on $B$ is non-monotonic. Figs. \ref{FIG1} and \ref{FIG2} show, for the equatorial plane,  how the boundary of $SLS$ depends on the magnetic field $B$. The analysis reveals that the magnetic field defines the geometry of the ergoregion.

When $B=0$ we recover the Reissner-Nordstr\"om metric, which does not possess $SLS$; when $0 < B \leq B_{1,2}\left(2M\right)$ the $SLS$ grows until it reaches its maximum size in $r_{SLS}^{\rm max}=2M$. Note that if $B= B_{1,2} \left( 2M \right)$, the maximum size of $SLS$ is independent of the BH charge $Q$. For $B_{1,2} \left( 2M \right)< B \leq B_{2,3} \left( r_+ \right)$, the $SLS$ decreases until it reaches its minimum in $r_{SLS}^{\rm min}= r_+$, when $B=B_{2,3} \left( r_{+} \right)$. For $B=B_{3,4}\left(2M\right)$ we find another maximum, and for $B_{3,4}\left(2M\right)<B$ $SLS$ tends asymptotically to $r_+$. This behavior has been qualitatively described in \cite{Nucamendi2022}, highlighting their numerical approximation, $B=0.4$, as the  magnetic field where the $SLS$ reaches its maximum in the equatorial plane. Our exact (analytical) result $B_{1,2}\left(2M\right)$ is consistent with theirs if $M=1$. It is worth mentioning that the non-monotonic behavior plays a crucial role in the energy extraction process, as the existence and size of the ergoregion determine the availability of negative energy trajectories.

\begin{figure*}[hbt]
\centering
\includegraphics[width=\textwidth]{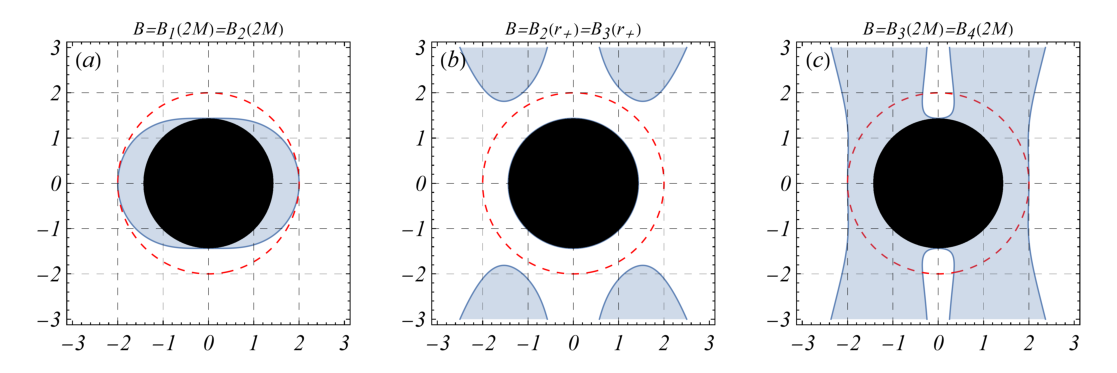}
\caption{\label{FIG3} The shaded areas are the ergoregions in a meridional plane $\left(\phi=\mathrm{constant}\right)$ for critical values of $B$, (a) $B=B_{1,2}\left(2M\right)=\left(\sqrt{2}-1\right)/M$, (b) $B=B_{2,3}\left(r_+\right)=2/r_+$; in this case the $SLS$ does not lie on the equatorial plane, but in  regions above and below it. (c) $B=B_{3,4}\left(2M\right)=\left(\sqrt{2}+1\right)/M$. The other parameters have been set to $M=1$ and $Q=0.9$. {The magnetic fields $B_{1,2}\left(2M\right)$ and $B_{3,4}\left(2M\right)$ share the same $SLS$ boundary, $r_{SLS}^{\rm max}=2M$}, denoted by the red dashed line.}
\end{figure*}

We can see that when $B=B_{2,3}\left(r_+\right)$, the radius of $SLS$ coincides with the event horizon $r_+$; recall that this is for the equatorial plane, and does not mean that the ergoregion disappears, but only that the ergosphere is  outside the equatorial plane. To elucidate this, in Fig. \ref{FIG3}  we  plot $SLS$ for a meridional plane ($\phi=$ constant) using Eq.\ (\ref{gtt}) for $B=B_{1,2}\left(2M\right),B_{2,3}\left(r_+\right),B_{3,4}\left(2M\right)$. When $B=B_{2,3}\left(r_+\right)$ we can observe that  $SLS$ is not situated in the equatorial plane, but in regions above and below it. Additionally, it is interesting that for $B_{1,2}\left(2M\right)$ and $B_{3,4}\left(2M\right)$ the behavior of  $SLS$ changes significantly outside the equatorial plane even though its radius is the same within it.

\section{Efficiency of the Penrose process and parameter-space characterization}\label{secEffi}

{Before proceeding, it is worth clarifying the meaning of the conserved energy used in the definition of the efficiency. In an asymptotically flat spacetime, the Killing energy associated with the vector $\xi^\mu=\partial_t$ can be normalized at spatial infinity and interpreted as the energy measured by inertial observers there \cite{Lasota2014}. The magnetized RN spacetime, however, is not asymptotically flat, and therefore such an ADM-type interpretation is not appropriate \cite{Brown1992}. Therefore, in the present case the quantity
\begin{equation}
    E_\xi=- \left(p_\mu+ q A_\mu\right) \xi^\mu,
\end{equation}
must be understood as the canonical conserved energy associated with the chosen stationary Killing field and with the electromagnetic gauge fixed by the magnetized solution \cite{Gibbons2013}. Since the same Killing field and the same gauge are used for all particles, the energy balance at the splitting point is well defined within this finite region description.}

{Consequently, the efficiencies discussed below should not be interpreted as efficiencies measured by asymptotic observers at flat infinity. Rather, they represent a Killing energy efficiency associated with the local splitting process in the magnetized background. Within this interpretation, energy extraction occurs when one fragment carries negative conserved Killing energy into the black hole, while the other fragment carries conserved Killing energy higher than that of the incident particle.}

In order to characterize the energy extraction process, we introduce the efficiency
\begin{equation}
     \eta=\frac{E_2-E_0}{E_0}=-\frac{E_1}{E_0},
\end{equation}
where $E_0$ is the energy of the incident particle, and $E_1<0$ corresponds to the fragment that falls into the BH. 

{It is important to emphasize that $\eta$ discussed in this Section is a local efficiency associated with the splitting event. The condition $\eta>0$ guarantees that, at the break-up point, the absorbed fragment can carry negative conserved Killing energy, while the other fragment carries  higher energy than the incident particle. However, this condition does not determine the subsequent radial evolution of the outgoing fragment. Escape must be checked independently from the effective potential of particle 2, requiring that no outer turning point prevents it from reaching the exterior region.}

{In what follows, we use $\eta$ to characterize the local extraction domain and to determine how the external magnetic field acts as a control parameter of the process. The corresponding escape supplement for neutral null fragments is discussed below in terms of the null effective potential.}

The condition $\eta>0$ requires the existence of negative energy states.
Using the general expression for the energies in Eq.\ (\ref{energies}), $\kappa =-1$ allows  $E_1<0$. To compare with \cite{Nucamendi2022}, we consider  a massive particle which decays into two massless particles,  $\tilde{\delta}_0= m_0^2$ and $\tilde{\delta}_1 =\tilde{\delta}_2=0$; the efficiency in this case is

\begin{equation}\label{efimagnetized}
    \eta=\frac{1}{2}\left(\frac{\sqrt{\left(E_0+q_0 A_t\right)^2+m_0^2g_{tt}}+\left(q_1-q_2\right)A_t}{E_0}-1\right).
\end{equation}
This expression shows explicitly that the efficiency is determined by two competing contributions: the gravitational term encoded in $g_{tt}$ and the electromagnetic interaction through the potential $A_t$. The interplay between these terms is responsible for the existence of negative energy states and therefore the viability of the extraction process that is determined by the condition $\eta\geq 0$, 
\begin{equation}\label{exregion}
\eta \geq 0 \quad \to  \quad   m_0^2 g_{tt}+4q_1 A_t \left(E_0 +q_2 A_t\right) \geq 0,
\end{equation}
where $\eta=0$ corresponds to the boundary of the extraction region; this condition defines the limits in parameter space beyond which negative energy states cease to exist. Eq. (\ref{exregion}) also shows that the extraction process is not solely determined by geometry (through $g_{tt}$), but also by the electromagnetic coupling between the BH and the test particle charges. To ensure the consistency between Eqs.\ (\ref{efimagnetized}) and (\ref{exregion}), it is necessary to impose additional constraints to avoid spurious solutions. In specific these conditions are,
\begin{equation}
   \left(E_0+q_0 A_t\right)^2+m_0^2 g_{tt}>0, \quad E_0 -\left(q_1-q_2\right)A_t>0.
\end{equation}
These conditions restrict the physically admisible region of the parameter space.

For chargeless particles ($q_i=0$), Eq.\ (\ref{exregion}) reduces to $g_{tt} \geq 0$ recovering the results discussed in the previous subsection, comprised in Eq.\ (\ref{Bcondition2}).

In magnetized RN BH, both $g_{tt}$ and $A_t$ depend explicitly on the magnetic field $B$  and, consequently, the efficiency $\eta$ becomes a nontrivial function of $B$. Taking the derivative of Eq.\ (\ref{efimagnetized}) with respect to $B$, we obtain the critical points of the efficiency from,
\begin{equation}\label{criticBeta}
\begin{aligned}
&\left(2q_0\left(E_0+q_0 A_t\right)A_{t,B}+m_0^2 g_{tt,B}\right)^2-4\left(q_1-q_2\right)^2\\
&\qquad\times A_{t,B}^2\left(\left(E_0+q_0 A_t\right)^2+m_0^2 g_{tt} \right)=0,
\end{aligned}
\end{equation}
where subindex $B$, denotes partial derivative with respect to $B$ and determines the extrema of efficiency for a given break-up point. Eq.\ (\ref{criticBeta}) reveals that the magnetic field affects the efficiency through two distinct channels: ($i$) by modifying the geometry via $g_{tt}$, and ($ii$) by changing the electromagnetic potential $A_t$. The competition between these effects leads to the appearance of critical values of $B$ for which efficiency is extreme.

The solutions for $B$  of  Eq. (\ref{criticBeta}) correspond to critical points of efficiency for a given break up point. In \cite{Nucamendi2022}, the efficiency $\eta$ of the process is analyzed in the vicinity of the event horizon $r_+$ for two particular cases;  firstly  for  chargeless particles, $q_0=q_1=q_2=0$. The second case considered is for $q_0=0$ and $q_1+q_2=0$. Both cases will be revisited below, presenting the exact analytical expressions.

\subsection{Case I: $q_0=q_1=q_2=0$}
\begin{figure*}[htb]
\includegraphics[width=8cm,height=5.5cm]{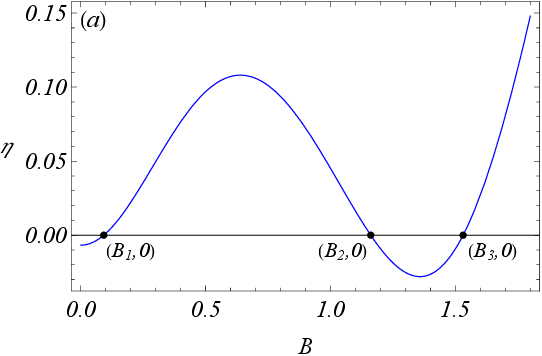}\quad \includegraphics[width=8cm,height=5.5cm]{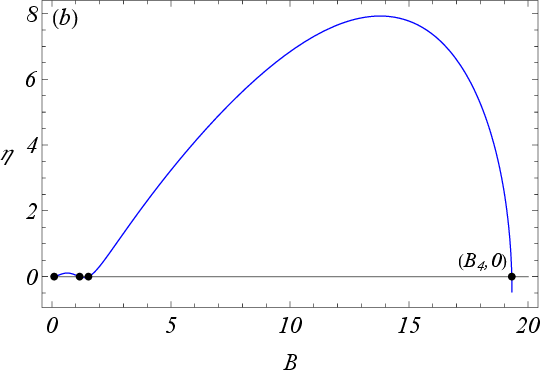}
\caption{\label{FIG4} It is illustrated the efficiency as a function of $B$ for chargeless particles, in the equatorial plane and for $M=1$ and $Q=0.9$. The event horizon is located at $r_+= M+\sqrt{M^2-Q^2}= 1.4359 $. The break up point has been set as $r_*=1.5$ such that $r_* >r_+$. The efficiency is a non-monotonic function, and exhibits two regions in the magnetic field parameter space where the efficiency is positive, and it is negative in the intervals $B < B_1\left(r\right)$, $B_{2}\left(r\right)< B < B_3\left(r\right)$ and $B > B_4\left(r\right)$. The zeroes of $\eta$ are the black dots. (b) the same than in (a) for a wider range of $B$,  note that energy  extraction is not possible for $B > B_4$. }
\end{figure*}

\noi Considering the magnetized RN BH, the efficiency $\eta$ for chargeless particles takes the form,
{\small
\begin{equation}\label{rnmefi0}
    \eta= \frac{1}{2}\left(\sqrt{1+4B^2Q^2\left(\frac{4-B^2 r_*^2}{4+B^2 r_*^2}\right)^2-\frac{\bar{F}\left(4+B^2 r_*^2\right)^2}{16}}-1\right),
\end{equation}
}
where $\bar{F}=F(r_*)$, where $F\left(r\right)$  is the metric function in Eq.\ (\ref{rnmmetric2}).
In the neutral case ($q_0=q_1=q_2=0$), the condition for energy extraction reduces to the requirement $g_{tt}>0$, i.e. the sign of $g_{tt}$ directly determines the existence of negative energy states. In regions where $g_{tt}<0$ energy extraction is not possible. The umbral or threshold  values of $B$ that allow $\eta>0$, for a given break-up point, are
\begin{equation} 
\begin{aligned}
     &B_1\left(r\right) < B < B_2\left(r\right),\\
     &B_3\left(r\right) < B < B_4\left(r\right),
\end{aligned}
\end{equation} 
where $B_i\left(r\right)$ are given in Table\ \ref{Tab1}. In Fig.\ \ref{FIG4} are displayed the two regions in the magnetic field parameter space where the extraction is possible for a break-up point $r_*>r_+$, where the minimum magnetic field  required for the efficiency to be positive is $B=B_1\left(r\right)$. Although the efficiency exhibits local maxima and minima, determining the corresponding values of $B$ for an arbitrary break-up point is quite complicated; however, they can be obtained analytically at the event horizon.

\begin{figure}[htb]
\includegraphics[width=8cm,height=5.5cm]{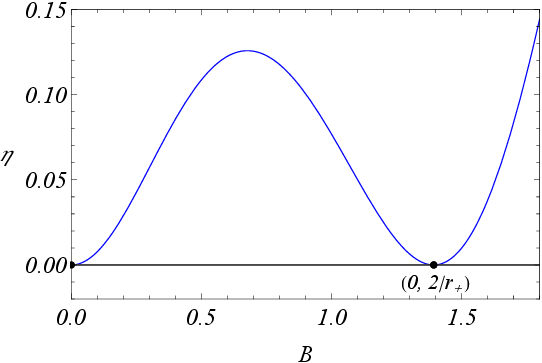}
\caption{\label{FIG5} It is illustrated the efficiency for chargeless particles as a function of $B$ with $r_* \to r_+$, in the equatorial plane and  for $M=1$ and $Q=0.9$. At $B=B^{\mathrm{max}}\left(r_+\right)$ the efficiency has a local maximum.}
\end{figure}

As we have mentioned, the extraction region for uncharged test particles is delimited by an inner and an outer radii, which are the event horizon $r_+$ and $SLS$, respectively. When the break up point approaches the inner limit ($r_* \to r_+$)  the umbral magnetic fields are as follows: $B_1\left(r\right) \to 0$, $B_{2,3}\left(r\right) \to 2/r_+$ and $B_4\left(r\right) \to \infty$ (see Table\ \ref{Tab2}). For these limiting cases, the efficiency is illustrated in Fig.\ \ref{FIG5}, where there are two degenerate solutions for $\eta=0$. In addition, note that for $r_* \to r_+$ the efficiency  is always positive, excluding the value $B_{2,3} \left( r_+ \right)$ where it vanishes. \\

\noi To compare the results presented in \cite{Nucamendi2022} with ours and to obtain an analytic expression of the efficiency for $r_* \to r_+$, we evaluate Eq.\ (\ref{rnmefi0}) in the vicinity of $r_+$, this yields
\begin{equation}\label{rnmefi1}
    \eta= \frac{1}{2}\left(\sqrt{1+4B^2Q^2\left(\frac{4-B^2 r_+^2}{4+B^2 r_+^2}\right)^2}-1\right),
\end{equation}
where $r_+=M+\sqrt{M^2-Q^2}$. It is important to emphasize that Eq.\ (\ref{rnmefi1}) has a correcting term compared to the one presented in \cite{Nucamendi2022}; this suggests a typographical omission\footnote{In  \cite{Nucamendi2022}, in the metric function $g_{tt}=-F H+H^{-1}r^2\omega \sin^2\theta$, and the correct expression is $g_{tt}=-F H+H^{-1}r^2\omega^2 \sin^2\theta$. With their equation for $g_{tt}$ into Eq.\ (\ref{efimagnetized}) its efficiency is recovered.} in \cite{Nucamendi2022}.

\begin{figure}[htb]
\includegraphics[width=8cm,height=5.5cm]{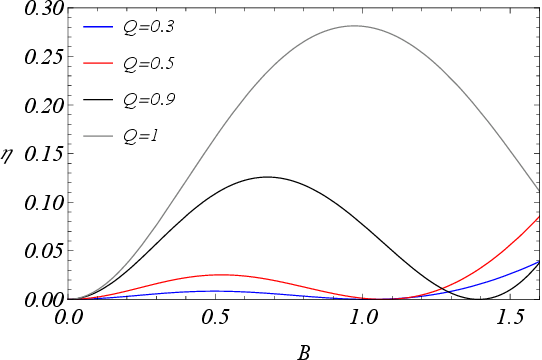}
\caption{\label{FIG6} Efficiency for non charged particles as a function of $B$ in the equatorial plane,  for $M=1$ and different values of BH charge $Q$. For $B<B^\mathrm{max}\left(r_+\right)$ the efficiency grows, otherwise the efficiency decreases. In $B=B^\mathrm{max}\left(r_+\right)$ the efficiency has a local maximum.}
\end{figure}

Fig.\ \ref{FIG6} presents the efficiency of the extraction process as a function of the magnetic field for different values of the BH charge, $Q$. In contrast to the case in Fig. \ref{FIG4} for $r_* >r_+$ the extraction threshold exhibits degenerate behavior. In this case, the condition $\eta > 0 $ is  satisfied for any $B > 0$, i.e. for any strenght of the magnetic field energy extraction is allowed. This is consistent with the fact that the horizon $r_{+}$ is the lower limit of the generalized ergoregion.

The  efficiency  behavior for $r_* \to r_+$ is clearly non-monotonic, exhibiting a local maximum at critical value $B=B^\mathrm{max}\left(r_+\right)$ and vanishing at $B=B^\mathrm{min}\left(r_+\right)$. To determine the critical values of $B$, using Eq.\ (\ref{criticBeta}) and $q_0=q_1=q_2=0$, reduces to the condition
\begin{equation}
    \frac{\partial g_{tt}}{\partial B}=0,
\end{equation}
whose solutions for $B$ can be obtained analytically as 
\begin{equation}
    B^\mathrm{max}\left(r_+\right)=\frac{2\sqrt{\sqrt{5}-2}}{r_+}, \quad B^\mathrm{min}\left(r_+\right) =\frac{2}{r_+}.
\end{equation}
Note that $B=B^\mathrm{min}\left(r_+\right)$ is directly related to $B_{3,4}\left(r_+\right)$ in Table \ref{Tab2}. When $B=B^\mathrm{max}\left(r_+\right)$, the efficiency reaches its local maximum, explicitly,
\begin{equation}
    \eta\left(r_+,B^\mathrm{max}\right)=\frac{1}{2}\left(\sqrt{1-8\left(11-5\sqrt{5}\right)\frac{Q^2}{r^2_+}}-1\right).
\end{equation}
In contrast, the vanishing efficiency at $B=B^\mathrm{min}$ is associated with the collapse of the ergoregion, when $SLS$ coincides with the event horizon in the equatorial plane. This result demonstrates that the magnetic field plays a dual role: it can both enhancing or inhibiting the extraction process, depending on its magnitude. 

In the case of chargeless particles, the extraction domain is not simply connected in the magnetic field parameter space, consisting of regions above and below the equatorial plane, see Fig.\ \ref{FIG3}b. For each fixed radius outside the horizon, two disconnected windows of allowed extraction arise, separated by a forbidden band. These windows merge on the horizon.

As we have mentioned, the ergoregion is the region where energy can be extracted and is bounded by the BH horizon and the $SLS$ which for certain values of magnetic field (see Table\ \ref{Tab3}), reaches its maximum size in the equatorial plane $r_{SLS}^{\rm max}=2M$. Since we have discussed what happens to efficiency for $r_* \to r_+$, it is natural to ask what happens when the break-up  point occurs at $r_* \to2M$. Unlike the horizon case, where the efficiency develops a local maximum, at $r_* =2M$ the efficiency remains non-positive and exhibits a finite minimum where the magnetic field and efficiency are,

\begin{equation}
\begin{aligned}
    & \eta\left(2M, B^{\mathrm{min}}\right)= \frac{1}{2}\left(\sqrt{1-2\left(\frac{2\left(\sqrt{2}-1\right)Q}{M}\right)^2}-1\right),\\
    & B^{\mathrm{min}}\left(2M\right)=\frac{\sqrt{2\sqrt{2}-1}}{M}.
\end{aligned}
\end{equation}

\begin{figure}[htb]
\includegraphics[width=8cm,height=5.5cm]{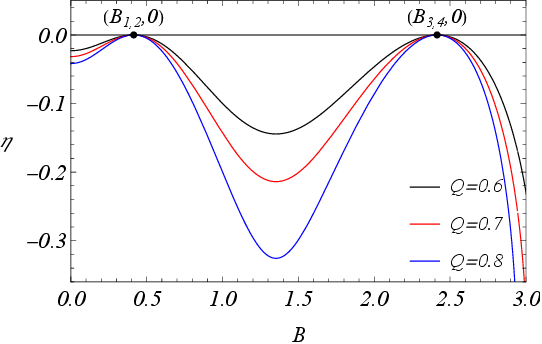}
\caption{\label{FIG7} It is illustrated that for break up points $r_* \geq 2M$ 
energy extraction is not possible because the efficiency is negative. The efficiency for non charged particles as a function of $B$ in the equatorial plane for $r_* \to 2M$, for $M=1$ and different values of BH charge $Q$ is always negative and exhibits two maxima, $\eta=0$, for the magnetic fields $B_{1,2}\left(2M\right)=\left(\sqrt{2}-1\right)/M$ and $B_{3,4}\left(2M\right)=\left(\sqrt{2}+1\right)/M$. The minimum local occurs at $B^\mathrm{min}\left(2M\right)=\sqrt{2\sqrt{2}-1}/M$. }
\end{figure}

In this limiting case and for  $r > 2M$, energy extraction is not possible, as the maximum efficiency is zero, as shown in Fig.\ \ref{FIG7}. Subsequently in case $q_i \neq 0$, we will show that if the particles are charged the extraction region extends beyond $r=2M$.

\begin{figure}[htb]
\includegraphics[width=8cm,height=5.5cm]{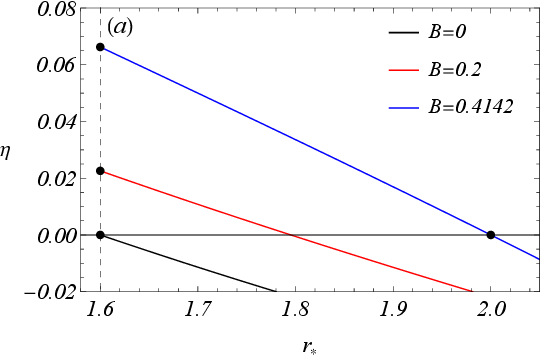}
\includegraphics[width=8cm,height=5.5cm]{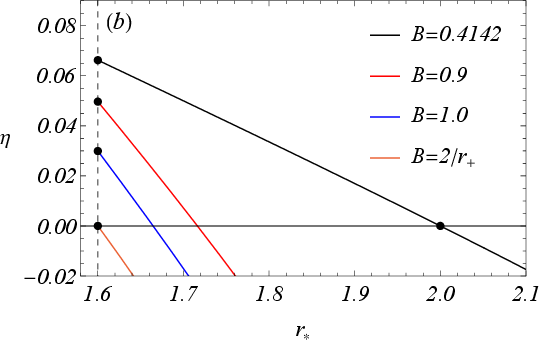}
\caption{\label{FIG700}  Efficiency for neutral particles as a function of the break-up point $r_*$ in the equatorial plane for several values of the magnetic field $B$ in the magnetic $0 \leq B  \leq 2/r_+$, in particular (a) corresponds to the domain of magnetic field $0\leq B\leq B_{1,2}\left(2M\right)$ and (b) corresponds to the domain $B_{1,2}\left(2M\right)\leq B\leq 2/r_+$, in both case we have considered $M=1$ and $Q=0.8$. Each curve corresponds to the efficiency of the process, the condition $\eta>0$ denotes the extraction region, and the crosses by zero $\eta=0$ indicates the boundary of it. The magnetic field $B=B_{1,2}\left(2M\right)$ produces the largest extraction region, extending up to $r_{SLS}^\mathrm{max}=2M$.}
\end{figure}

To further clarify the role of the break-up point in the neutral extraction process, it is useful to analyze the efficiency as a function of the radial parameter $r_*$ for several representative magnetic-field values. The behavior of the efficiency for magnetic field regime  $0\leq B\leq 2/r_+$ is illustrated in Fig.\ \ref{FIG700}. In this regime, the efficiency decreases monotonically as the break-up point moves away from the horizon, vanishing when the outer boundary of the extraction region is reached. For $B=0$, the efficiency vanishes identically, consistent with the absence of ergoregion in RN BH. A qualitatively distinct behavior appears around $B=B_{1,2}\left(2M\right)$ (see Table\ \ref{Tab3}), which corresponds to one of the configurations where the $SLS$ reaches its maximum extension in the equatorial plane. For $B<B_{1,2}\left(2M\right)$, increasing the magnetic field enlarges the extraction region and enhances the efficiency, whereas for $B>B_{1,2}\left(2M\right)$, the efficiency begins to decrease and the extraction region shrinks toward the event horizon. This transition reflects the non-monotonic dependence of the ergoregion on the magnetic field.

\begin{figure}[htb]
\includegraphics[width=8cm,height=5.6cm]{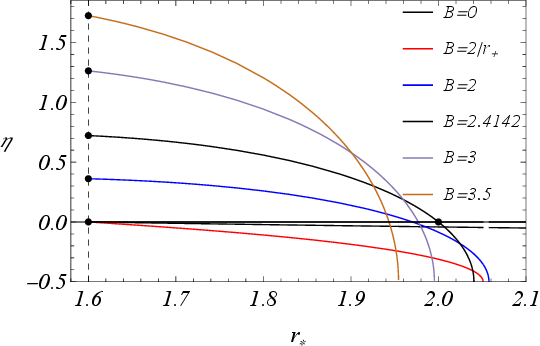}
\caption{\label{FIG7_1} Efficiency for neutral particles as a function of the break-up point $r_*$ in the equatorial plane for several values of the magnetic field $B$ in the magnetic $2/r_+ \leq B  \leq \infty$, with $M=1$ and $Q=0.8$. Each curve corresponds to the efficiency of the process, the condition $\eta>0$ denotes the extraction region, and the crosses by zero indicate the boundary of it. The magnetic field $B=B_{3,4}\left(2M\right)$ produces the largest extraction region, extending up to $r_{SLS}^\mathrm{max}=2M$.}
\end{figure}

In Fig.\ \ref{FIG7_1}, we display the behavior of the efficiency for a magnetic field in regime $2/r_+\leq B \leq \infty$. As the magnetic field increases, a finite extraction region emerges and extends outward from the horizon. The largest extraction region is obtained for a magnetic field, $B=B_{3,4}\left(2M\right)$ (see Table\ \ref{Tab3}), for which the $SLS$ reaches its maximum radius $r_{SLS}^\mathrm{max}=2M$. This behavior is shown in Fig.\ \ref{FIG7_1}. For values of magnetic field  $B>B_{3,4}\left(2M\right)$, the extraction region starts to shrink again, and the point where $\eta=0$ moves toward the event horizon. In particular, when $B=2/r_+$, the efficiency vanishes exactly at the horizon, signaling the collapse of the equatorial ergoregion. Beyond this value, the extraction region reappears with a progressively smaller radial extension, in agreement with the non-monotonic behavior of the $SLS$ discussed previously.

It worth noticing that the shape of the efficiency curves changes qualitatively according to the magnitude of the magnetic field. For $B<2/r_+$, the efficiency decreases almost linearly between the event horizon and the outer boundary of extraction region, see Fig.\ \ref{FIG700}. In contrast, for larger magnetic fields the curves become increasingly distorted and the extraction region shrinks toward the horizon, see Fig.\ \ref{FIG7_1}. This behavior reflects the non trivial dependence of the generalized ergoregion on the magnetic field and illustrates how the magnetic field simultaneously controls both the size of the extraction region and the magnitude of the efficiency.

\subsection{Case II: $q_0\neq 0$, $q_1\neq 0$ and $q_2\neq 0$}
The presence of electromagnetic interaction between the BH and charged test particles introduces qualitatively new features in the energy extraction process. In contrast to the neutral case, where extraction is purely geometric and determined by the sign of $g_{tt}$, the charged case leads to a coupled geometric and electromagnetic criterion.\\ 

To determine the condition that governs the boundary of the extraction region it is necessary to impose $\eta=0$, that is, equality in Eq.\ (\ref{exregion}),
\begin{equation}\label{etazero}
m_0^2 g_{tt}+4q_1 A_t \left(E_0 +q_2 A_t\right) = 0.
\end{equation}
Recalling that $q_1$, $q_2$ are the charges of the falling and escaping particles, respectively; the previous equation generalizes the stationary condition $g_{tt}=0$ and defines an effective energy extraction radius; also shows explicitly that the existence of negative energy states is governed by the interplay between gravitational and electromagnetic interaction. 

\begin{figure}[htb]
\includegraphics[width=8cm,height=7.5cm]{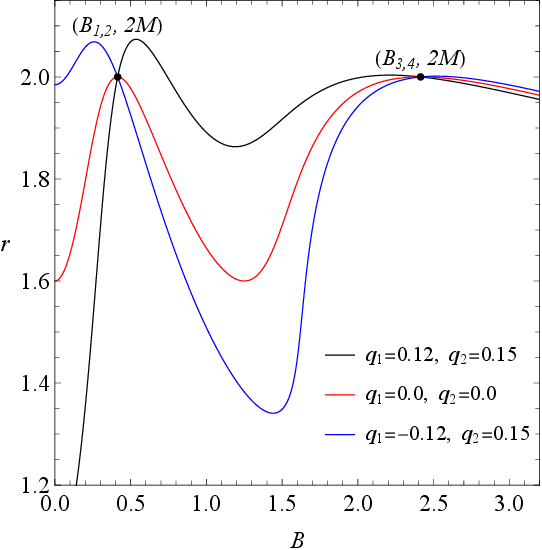}
\caption{\label{FIG8} Comparison of the radii boundary limits of the extraction region for charged particles $q_1$ and $q_2$ (charges of the falling and escaping particles, respectively) as a function of the magnetic field $B$ (black and blue graphics) and the static limit surface ($SLS$) for uncharged particles (red curve). Depending on the magnetic field the radius of the extraction region may be larger or smaller as compared with the $SLS$, $r_{\rm eg } > r_{SLS}$ or $r_{\rm eg } < r_{SLS}$; in all cases $r_{\rm eg }$ is the same at the extrema of $SLS$ (black dots), that are the $B$ satisfying $g_{tt}=0$ for uncharged particles.}
\end{figure}

Fig. \ref{FIG8} displays solutions of Eq. (\ref{etazero}) for $r$ in terms of $B$ where we can noticed that for charged case, the extraction region no longer coincides necessarily with the geometrically defined ergoregion, $g_{tt}=0$. For charged particles, the boundary of the extraction region reaches radii larger than $SLS$, $r_{\rm eg } > r_{SLS}$; or smaller than the $SLS$, $r_{\rm eg }<r_{SLS}$, and  the magnetic field strength is one of the parameters that modulates or distinguishes these scenarios. Note that these differences $r_{\rm eg }>r_{SLS}$ or $r_{\rm eg }<r_{SLS}$ are strongly attached to the maxima in $g_{tt}=0$. Another important point is that the extraction region persists, regardless of the sign of the charges of the particles, even if  $q_1$ and $Q$ are of the same sign. This differs significantly from RN BH  where the necessary condition for  extraction to occur is $q_1 Q<0$.\\

 Substituting the explicit expressions for $g_{tt}$ and $A_t$ into  Eq. (\ref{etazero}) leads to a bi-quartic equation in the magnetic field $B$,
 \begin{equation}
     B^8-a_0 B^6 + a_1 B^4 - a_2 B^2 +a_3 =0,
 \end{equation}
 whose coefficients depend on the BH parameters and on the properties of the particles. Although this equation can be solved analytically, its structure obscures the physical interpretation. To make progress the following parameterization is introduced,
 
 \begin{equation}\label{Banzats2}
    B\left(r, Q, x\right) = \frac{2}{r}\sqrt{x}
\end{equation}
which reduces the problem to a quartic equation, 
\begin{equation}\label{a0charge}
 x^4- c_0 x^3+ c_1 x^2-c_2 x+c_3=0,
\end{equation}
where
\begin{equation}
\begin{aligned}
   &c_0=\frac{4\left(\kappa_1-\kappa_0 - \kappa_2 \right)}{\kappa_0}, \quad c_1= \frac{2\left(4\kappa_1+3\kappa_0-10 \kappa_2\right)}{\kappa_0},\\
   & \qquad c_2= \frac{4\left(\kappa_1-\kappa_0+5\kappa_2\right)}{\kappa_0}, \quad c_3=\frac{\kappa_0 + 4 \kappa_2}{\kappa_0},\\
   & \quad \kappa_0 = m_0^2 \Delta-4Q^2 q_1 q_2\quad \kappa_1 = 4Q^2\left(m_0^2-4 q_1 q_2\right)\\
   & \qquad\quad \kappa_2 =E_0 Q q_1 r,\quad  \Delta=r^2-2M r+Q^2.
\end{aligned}
\end{equation}
Eq.\ (\ref{a0charge}) admits up to four real solutions, which we denote as $x_i$, leading to four corresponding values for the magnetic field, $B_i \left(r\right)=B\left(r, Q, x_i\right)$. These solutions define the boundaries of the extraction region in parameter space, the explicit forms of $x_i$ are\\ 
\begin{equation}\label{xsol}
    \begin{aligned}
        &x_{1}= \frac{c_0}{4}-{\alpha}-\sqrt{\alpha^2-\beta_2}, \quad x_{2}=\frac{c_0}{4}-{\alpha} +\sqrt{\alpha^2-\beta_2},\\
        &x_{3}= \frac{c_0}{4}+{\alpha} -\sqrt{\alpha^2-\beta_1}, \quad x_{4}=\frac{c_0}{4}+{\alpha} +\sqrt{\alpha^2-\beta_1},
    \end{aligned}
\end{equation}
where,
\begin{equation}
    \begin{aligned}
        & \beta_{1,2} = 2\alpha^2 - 3 A_2 +2\epsilon_{1,2} \alpha^{-1} A_1, \quad \epsilon_1 = -1, \quad \epsilon_2=1,\\
        &\alpha = A_2+2\sqrt{A_0}\cos\left(\frac{1}{3}\arccos{\left(\frac{B_0}{A_0^{3/2}}\right)}\right), \quad A_0=A_2^2-A_3,\\
        &\quad B_0 = \frac{2A_2^3+A_1^2-3A_2 A_3}{2},\quad A_1=\frac{c_0^3-4c_0 c_1+8c_2}{64},\\ 
        &A_2 = \frac{c_0^2}{16}-\frac{c_1}{6}, \quad A_3 = \frac{4\left(2c_1-c_0^2\right)^2-c_0^4+16c_0 c_2-64 c_3}{768}.
    \end{aligned}
\end{equation}
Although  explicit solutions are algebraically complex, their physical role is to define the boundaries of the extraction region. A key requirement for physical relevance is that the corresponding values of $B_i\left(r\right)$ be real and positive. This imposes constraints on the allowed parameter space and naturally leads to an ordering when four real solutions exist,
\begin{equation}
    B_1 \left(r\right)\leq B_2 \left(r\right)\leq B_3 \left(r\right)\leq B_4\left(r\right),
\end{equation}
which generalizes the chargeless structure. In the limit $q_1 \to 0$ and $q_2\to 0$, the coefficients reduce to their geometric counterparts and the solutions recover the critical fields discussed in Table\ \ref{Tab1}-\ref{Tab3}.

\begin{figure}[htb]
\includegraphics[width=8cm,height=5.5cm]{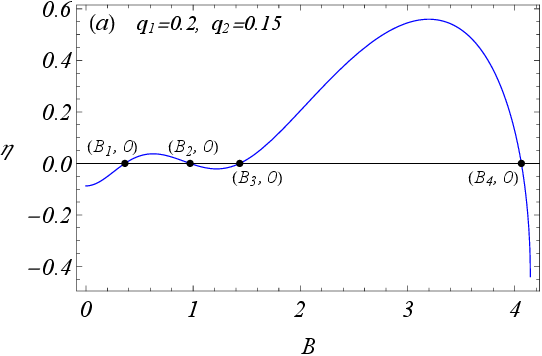}\\
\includegraphics[width=8cm,height=5.5cm]{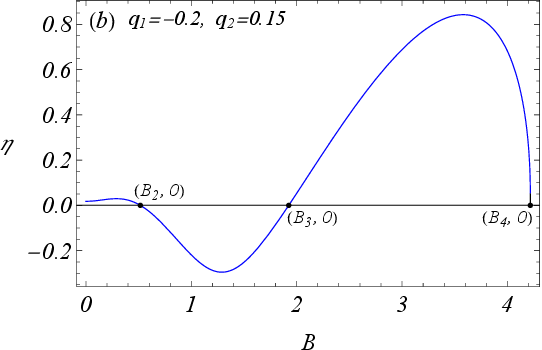}
\caption{\label{FIG10} It is illustrated the extraction efficiency for charged particles as a function of $B$ in the equatorial plane, for $M=1$ and $Q=0.8$. The event horizon is located at $r_+= M+\sqrt{M^2-Q^2}= 1.6 $. The break up point has been set as $r_*=1.9$ such that $r_*>r_+$. (a) For $q_1=0.12$ and $q_2=0.15$, the efficiency two disconnected intervals of positive efficiency separated by a forbidden band where the extraction is not allowed. (b) For $q_1=-0.12$ and $q_2=0.15$, one of the region with negative efficiency disappears as a consequence of non real solutions for $B$ according to the condition Eq.\ (\ref{qthr0}).}
\end{figure}

When four real solutions exists, the efficiency exhibits a non-monotonic dependence on the magnetic field, leading to two disconnected intervals of positive efficiency separated by a forbidden region where negative energy states are not allowed. This behavior is shown in Fig.\ \ref{FIG10}a.  

The structure of the solutions changes qualitatively depending on the relation between the particle charges. In particular, the inequality
\begin{equation}\label{qthr0}
    q_1 > \frac{m_0^2\left(r^2-2M r+Q^2\right)}{4Q\left(Q q_2-E_0 r\right)},
\end{equation} 
provides a necessary (though not sufficient) condition for the existence of four real solutions. When this condition is reversed, the number of physical solutions decreases and the structure of the extraction region is modified as illustrated in Fig.\ \ref{FIG10}a.

Unlike the neutral case, where energy extraction requires a nonzero magnetic field, the charged case admits configurations in which the extraction is allowed for $B=0$; see the efficiency in  Fig.\ \ref{FIG10}b for this magnetic field strength. This occurs when the electromagnetic interaction in itself is sufficient to generate negative energy states.

\subsubsection{Boundary conditions and $B$ as a control parameter}

The transition between disconnected and connected extraction regions can be characterized by the extrema of the magnetic field.  Eq. (\ref{etazero}) and its implicit derivative with respect to $B$ (imposing $d r/d B = 0$),  constitute the system of equations 
\begin{equation}
    \begin{aligned}
        &m_0^2 g_{tt}+4q_1 A_t \left(E_0 +q_2 A_t\right) = 0\\
        & m_0^2 g_{tt,B}+ 4q_1 A_{t,B}\left(E_0+ 2q_2 A_t\right)=0,
    \end{aligned}
\end{equation}
where the sub-index $B$ denotes partial derivative with respect to $B$. Solving these equations yields the following charge configurations
\begin{equation}\label{chargesmin}
    \begin{aligned}
        &q_1^{\mathrm{crit}} = \frac{m_0^2\left(A_t g_{tt,B}-2 A_{t,B} g_{tt}\right)}{4 E_0 A_{t} A_{t,B}},\\
        &q_2^{\mathrm{crit}} =\frac{E_0 \left(A_{t,B} g_{tt} - A_{t} g_{tt,B}\right)}{A_t\left(A_t g_{tt,B}-2 A_{t,B} g_{tt}\right)},
    \end{aligned}
\end{equation}
that define the conditions under which the two disconnected extraction windows merge into a single continuous region; the degeneracy in the magnetic field solutions, $B_2\left(r\right)=B_3\left(r\right)$ removes the forbidden band and the efficiency becomes positive in the interval $B_1\left(r\right)<B<B_4\left(r\right)$, except at the degenerate point where $\eta=0$. In addition, these charge configurations allow us to  control the non-trivial local minimum ($B=0$) of $SLS$, restricting to scenarios where the radius of the extraction region is located outside the event horizon, $r> r_+$.

An important feature of the critical charge configurations for magnetized RN is the presence of a singular behavior in $q_2^{\mathrm{crit}}$ at the radius $r=2M$; this singular behavior can be made explicit by inspecting the denominator of Eq.\ (\ref{chargesmin}). In fact, after substituting the explicit expressions for $g_{tt}$ and $A_t$, the denominator of $q_2^{\mathrm{crit}}$ contains the  factor $\left(r-2M\right)$. Consequently, $q_2^{\mathrm{crit}} \sim 1/\left(r-2M\right)$. This divergence is not a spurious algebraic artifact. Rather, it reflects the special role of $r=2M$, which corresponds to the maximum equatorial extension of  $SLS$ in the neutral case. At this radius, the degeneracy condition $B_2\left(r\right)=B_3\left(r\right)$ cannot be maintained with a finite value of $q_2$. This signals a breakdown of the critical charge configuration at $r=2M$, where the transition between disconnected and connected extraction windows becomes singular.

\begin{figure}[htbp!]
\includegraphics[width=8cm,height=5.5cm]{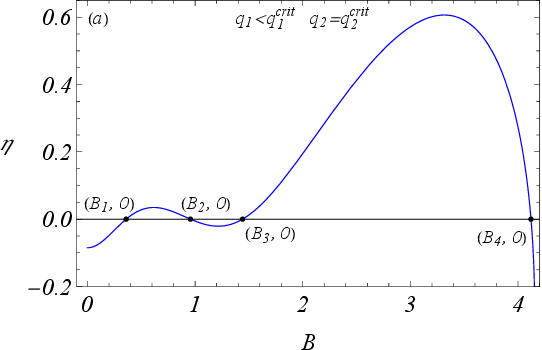}
\includegraphics[width=8cm,height=5.5cm]{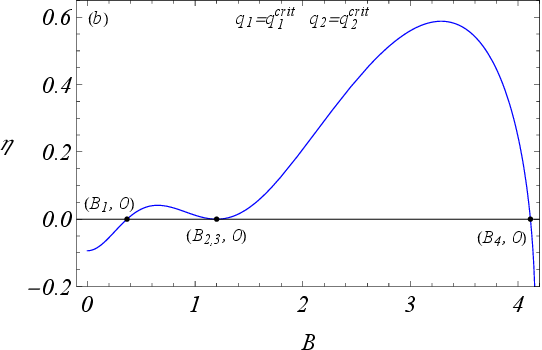}
\includegraphics[width=8cm,height=5.5cm]{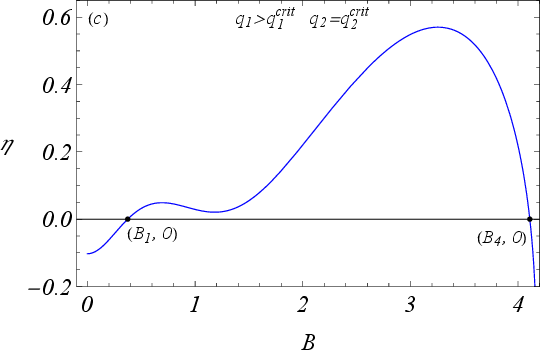}
\caption{\label{FIG12}  The efficiency as a function of the magnetic field for different charge configurations relative to the critical values $q_1^{\mathrm{crit}}$ and $q_2^{\mathrm{crit}}$, at the break up point $r_*=1.9$. The rest of parameters have been fixed as $M=1$ and $Q=0.8$. (a) For $q_1<q_1^{\mathrm{crit}}$, the extraction region ($\eta >0 $) consists of two disconnected intervals separated by a forbidden band. (b) At the critical configuration $q_1=q_1^{\mathrm{crit}}$, a degeneracy occurs $B_2\left(r\right)=B_3\left(r\right)$, and the forbidden band collapses to a single point. (c) For $q_1>q_1^{\mathrm{crit}}$, energy can be extracted in the interval $B_1\left(r\right) < B < B_4\left(r\right)$.}
\end{figure}

Fig.\ \ref{FIG12} illustrates the qualitative change in the structure of the extraction region as the particle charges approach the critical configurations defined by Eq.\ (\ref{chargesmin}). We have fixed the values of $q_2=q_2^{\mathrm{crit}}$ and identify three different scenarios. In Fig.\ \ref{FIG12}a,  $q_1<q_1^{\mathrm{crit}}$, the efficiency exhibits two disconnected intervals of positive values, separated by a forbidden band where the energy extraction is not allowed. As we have discussed, this behavior reflects the existence of four distinct real solutions for the magnetic field that define the boundaries on the extraction domain.

In Fig.\ \ref{FIG12}b is illustrated the critical configuration $q_1=q_1^{\mathrm{crit}}$, a degeneracy occurs in the solutions of the boundary equation, such that $B_2\left(r\right)=B_3\left(r\right)$. As a consequence, the forbidden band collapses to a single point and the two disconnected extraction windows merge.
In Fig.\ \ref{FIG12}c $q_1>q_1^{\mathrm{crit}}$ and the extraction regime is a single continuous interval, $B_1\left(r\right)<B<B_4\left(r\right)$, indicating that the efficiency remains positive throughout the entire range except at boundary points ($B_1\left(r\right)$ and $B_4\left(r\right)$). This transitions demonstrates that electromagnetic interaction does not merely shift the extraction thresholds but fundamentally reorganizes the structure of the allowed parameter space. Therefore, the particle charges act as control parameters that govern a transition between disconnected and connected extraction regimes, highlighting the non trivial interplay between gravitational and electromagnetic effects in magnetized BH spacetimes.

\begin{figure}[htbp!]
\includegraphics[width=8cm,height=5.5cm]{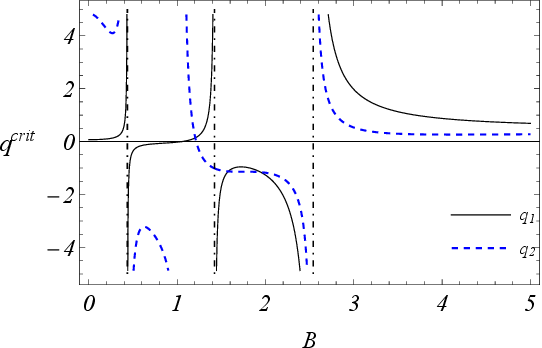}
\caption{\label{FIG13} Critical charge configurations $q_1^{\mathrm{crit}}$ and $q_2^{\mathrm{crit}}$ (black and dashed blue curve, respectively) as a function of the magnetic field $B$,obtained from Eq.\ (\ref{chargesmin}) {for a given break-up point, $r_\ast = 1.9$}. These charges determine the degeneracy condition $B_2\left(r\right)=B_3\left(r\right)$ that defines the transition disconnected--connected extraction regions. The graphics show that the transition is not defined by the signs of the charges, such that magnetized configurations remove the standard electrostatic constrain $q_1 Q < 0$, replacing it with a dynamical condition controlled by $B$. $q_1^{\mathrm{crit}}$ diverges at: $B= 2\left(\sqrt{2}-1\right)/r_*$, $B= 2\left(\sqrt{2}+1\right)/r_*$, $B=\left(2\sqrt{2\sqrt{2}-1}\right)/r_*$ and $q_2^{\mathrm{crit}}$: $B= 2\left(\sqrt{2}-1\right)/r_*$, $B= 2\left(\sqrt{2}+1\right)/r_*$, $B=2/r_*$, values indicated with the vertical lines.}
\end{figure}

An important feature of the critical charge configurations given by Eq.\ (\ref{chargesmin}) is that their sign is not fixed apriori. Depending on the interplay between the geometric contribution $g_{tt}$ and the electromagnetic potential $A_t$, the critical charges can take either positive or negative values. This implies that the transition between disconnected and connected extraction regions is not restricted to configurations where the particle charge is opposite to the BH charge. In particular, energy extraction can occur even when $q_1 Q>0$, in contrast with the standard RN BH case where the condition $q_1 Q<0$ is required \cite{DenardoRuffini1972}, replacing it with a dynamical condition controlled by $B$. Fig.\ \ref{FIG13} displays the critical charge configurations $q_1^{\mathrm{crit}}$ and $q_2^{\mathrm{crit}}$ as a function of the magnetic field $B$, obtained from Eq.\ (\ref{chargesmin}). The parameters are fixed as $M=1$, $Q=0.8$ and $r_*=1.9$. These curves determine the conditions under which the degeneracy $B_2\left(r\right)=B_3\left(r\right)$ occurs, where the change in sign of charge configurations is evident. The figure also reveals a nontrivial structure in the dependence on $B$, including divergences associated with the zeroes of the denominator in Eq.\ (\ref{chargesmin}). These singular points separate different regimes in parameter space, related to differences in the sign and magnitude of the critical charges.

 It is instructive to examine the weak-field behavior of the critical charge configurations. From Eq.\ (\ref{chargesmin}), in the limit $B\to 0$, one obtains

 \begin{equation}
 \begin{aligned}     
     & q_1^\mathrm{crit}\left(B\to 0\right)=\frac{4m_0^2\left(2M-r\right)}{7E_0 Q},\\
     &q_2^\mathrm{crit}\left(B\to 0\right)=\frac{E_0}{16}\left(\frac{9r}{Q}-\frac{7Q}{r-2M}\right).
 \end{aligned}
 \end{equation}
It is important to stress that these limiting expressions, do not reproduce the standard RN condition for energy extraction ($q_1 Q<0$). Rather, they should be interpreted as a remnant of the magnetized parameter-space structure. In particular, the behavior of $q_2^\mathrm{crit}$ in the limit $B\to 0$ is fully consistent with the general analysis presented above. The divergence ar $r=2M$ persists in this limit, confirming that is is not associated with a specific magnetic-field regime, but is instead a structural feature of the critical configuration.

Having established the weak-field behavior, we now consider the case in which the magnetic field becomes dominant. In the limit, $B\to \infty$, the critical charge configuration takes the asymptotic form
\begin{equation}
\begin{aligned}
& q_1^\mathrm{crit}\left(B\to \infty\right)=\frac{4m_0^2\left(2M-r\right)}{E_0 Q},\\
&q_2^\mathrm{crit}\left(B\to \infty\right)=\frac{E_0}{16}\left(\frac{r^2-2M r+Q^2}{Q\left(2M-r\right)}\right).
\end{aligned}
\end{equation}
We can observe that the dependence of $q_1^{\mathrm{crit}}$ and $q_2^{\mathrm{crit}}$ on the factor $\left(r-2M\right)$ and $1/\left(r-2M\right)$ respectively, is preserved, confirming that it is a structural feature of the critical configurations rather than a property of specific magnetic-field regime. On the other hand, the expression for $q_2^{\mathrm{crit}}$ can be written in terms of the function $\Delta=r^2-2M r+Q^2$, which encodes the causal structure of the underlying RN geometry,  explicitly stating that the strong-field limit retains a direct imprint of the spacetime structure.

Taken together, the weak and strong field limits show that the qualitative features of the critical charge configurations are governed by geometric quantities associated with the underlying spacetime, in particular the radial function $\Delta = r^2-2M r+Q^2$ and the distinguished scale $r=2M$, while the magnetic field acts as a control parameter that modulates their quantitative behavior.

\subsubsection{Dependence of critical charges on the break-up point}
To further characterize the structure of the extraction region, we analyze configurations where the boundary defined by 
\begin{equation}
    m_0^2 g_{tt}+4q_1 A_t \left(E_0 +q_2 A_t\right) = 0,\nonumber
\end{equation}
becomes extrema with respect to the radial coordinate. This is achieved by imposing the conditions
\begin{equation}
    m_0^2 g_{tt,r}+ 4q_1 A_{t,r}\left(E_0+ 2q_2 A_t\right)=0,
\end{equation}
where the sub-index denotes the partial derivative with respect to $r$. The explicit solutions for $q_1$ and $q_2$ are,
\begin{equation}\label{chargesminr}
    \begin{aligned}
        &q_1^{\mathrm{crit}} = \frac{m_0^2\left(A_t g_{tt,r}-2 A_{t,r} g_{tt}\right)}{4 E_0 A_{t} A_{t,r}},\\
        &q_2^{\mathrm{crit}} =\frac{E_0 \left(A_{t,r} g_{tt} - A_{t} g_{tt,r}\right)}{A_t\left(A_t g_{tt,r}-2 A_{t,r} g_{tt}\right)}.
    \end{aligned}
\end{equation}
These equations identify a marginal configuration where the extraction boundary reaches an extrema as a function of the break-up point where $\eta=0$. In such cases, the efficiency fulfill $\eta \leq 0$ without changing sign.

\begin{figure}[htbp!]
\includegraphics[width=8.5cm,height=5.5cm]{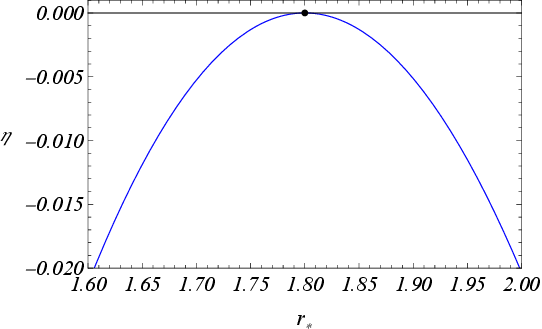}\\
\caption{\label{FIG14} Efficiency as a function of the break-up point $r_*$, evaluated at the critical charge configurations defined by Eq.\ (\ref{chargesminr}). The remaining parameters are fixed as $M=1$, $Q=0.8$, and $B=1$. The efficiency is negative and reaches $\eta=0$ only at  $r_*=1.8$ for critical charges $q_1^{\mathrm{crit}}=-3.202$ and $q_2^{\mathrm{crit}}=-1.303$, indicating that these configurations correspond to threshold states rather than optimal extraction.}
\end{figure}

As we illustrated in Fig.\ \ref{FIG14}, the efficiency evaluated at these critical configurations of charges is negative and vanishes only at isolated parameter values. These configurations correspond to marginal threshold points in the extended parameter space $\left(r, B, q_1, q_2\right)$ marking the transition between the extracting and non-extracting regimes rather than optimal efficiency states.

\begin{figure}[htbp!]
\includegraphics[width=8.5cm,height=5.5cm]{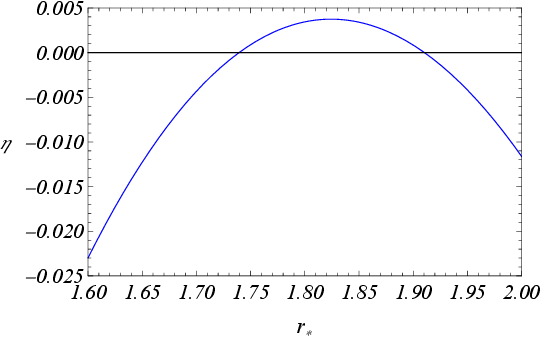}
\caption{\label{FIG15} Efficiency as a function of the break-up point $r_*$ with charges slightly different from critical charges in Eq.\ (\ref{chargesminr}) . The remaining parameters are fixed as $M=1$, $Q=0.8$, and $B=1$ and the charges has been set $q_1=-3.702$ and $q_2=-1.303$. The emergence of a positive efficiency interval shows that the critical charges mark the onset of extraction and separate non-extracting from extracting regimes.}
\end{figure}

However, small deviations from these critical charges lead to the emergence of positive efficiency, as shown in Fig. \ref{FIG15}. This behavior demonstrates that the critical configurations correspond to the threshold states where the boundary of the extraction region undergoes a transition.

It is worth emphasizing that once the full radial dependence of the efficiency is taken into account, the maximum efficiency is not necessarily attained at the event horizon. Instead, we find that the efficiency reaches its maximum at a finite radius $r_*>r_+$ where the interplay between the gravitational contribution encoded in $g_{tt}$ and the electromagnetic interaction through $A_t$ is optimized. This contrasts with analyses restricted to the near horizon region where the efficiency is typically evaluated in the limit $r_*\to r_+$. Our results show that such an approximation captures the local behavior of the process but does not, in general, determine its global maximum. Therefore,  the location of maximal efficiency is controlled by the competition between gravitational and electromagnetic contributions rather than by the horizon geometry alone. This highlights the importance of treating the break-up point as a dynamical parameter rather than fixing it at the horizon.

\subsection{Discussion: local extraction versus escape conditions}
To provide a more complete analysis of the subsequent evolution of the escaping particle (particle 2 in our analysis), we examine the necessary conditions under which such escape is possible, in particular, for non-charged fragments.

{Since in the present treatment the daughter escaping particle follows  a null trajectory, we consider the effective potential Eq.\ \eqref{potefectivo} with $\delta_2=0$  and $e_2=0$, this yields
\begin{equation}
    V_{\pm 2}=
    l_2\frac{-g_{t\phi}\pm\sqrt{g_{t\phi}^{2}-g_{tt}g_{\phi\phi}}}
     {g_{\phi\phi}}.
\end{equation}
Null circular geodesics, namely, photon orbits in the neutral case, are then selected with the additional extremal condition
\begin{equation}
\frac{dV_{+2}}{dr}=0.
\end{equation}
For the magnetized Reissner-Nordstr\"om spacetime this condition gives
\begin{equation}\label{photoregion}
\begin{aligned}
&(3r^2-5Mr+2Q^2)r^2B^2-4(r^2-3Mr+2Q^2)\\
&+8BQ r\sqrt{r^2-2Mr+Q^2}=0.
\end{aligned}
\end{equation}
Whose roots for the magnetic field are,
\begin{equation}
\begin{aligned}
    &\qquad\qquad B_{\rm ph}\left(r\right)=\frac{2}{r}\left(\frac{A_1}{2Q\sqrt{A_0}+\epsilon\sqrt{A_2}}\right),
\end{aligned}
\end{equation}
where
\begin{equation}
    \begin{aligned}
    &A_0=r^2-2M r+Q^2,\quad A_1=r^2-3M r+2Q^2,\\
    &A_2=\left(3r^2-5M r+2Q^2\right)A_1+4Q^2 A_0, \quad \epsilon= \pm 1.
    \end{aligned}
\end{equation}
The existence of a real photon region requires
\begin{equation}
    A_2\geq 0 \to  \left(3r^2-5M r+2Q^2\right)A_1+4Q^2 A_0\geq 0.
\end{equation}
This condition is the positivity of the discriminant of the quadratic Eq. \eqref{photoregion} for $B$. Thus, for fixed $r$, it determines the external magnetic field that corresponds to a radius in the photon region.}

{In the absence of the external magnetic field, $B=0$, one recovers the usual Reissner-Nordstr\"om photon region radius
\begin{equation}
    r_{\rm ph}^{\rm RN}=\frac{3M+\sqrt{9M^2-8Q^2}}{2},
\end{equation}
which satisfies $2M\leq r_{\rm ph}^{\rm RN}\leq 3M$ for $0\leq |Q|\leq M$.}

{Therefore, for an outward-directed massless fragment, a conservative escape criterion is
\begin{equation}\label{lowerbound}
    r_{\rm ph}\leq r_\ast,
\end{equation}
where $r_\ast$ is the break-up radius and $r_{\rm ph}$ denotes the photon region radius.}

{In the neutral case, the local extraction region is bounded by the event horizon and the $SLS$. Hence, local extraction and escape can be simultaneously associated with a massless fragment only if the break-up radius lies in the radial window
\begin{equation}\label{radialwindow}
r_{\rm ph}< r_\ast\leq r_{SLS}.
\end{equation}
This condition should be understood as a kinematical supplement to the local extraction condition $\eta>0$.}

{This observation motivates the analysis of two limiting configurations. First, the case where the photon region coincides with the event horizon, $r_{\rm ph}=r_+$. This would correspond to the least restrictive escape condition, since any outward-directed massless fragment produced outside the horizon and inside the extraction region would lie outside the photon region threshold. Evaluating  Eq. \eqref{photoregion} at the horizon gives
\begin{equation}
r_+(r_+-M)\left(r_+^2B^2+4\right)=0,
\end{equation}
that corresponds to a degenerate limiting configuration in the extreme case 
\begin{equation}
    r_+=M,\qquad |Q|=M.
\end{equation}
In the extreme limit $r=M$ and $Q^2=M^2$ (where $Q=\pm M$), Eq. \eqref{photoregion} leads the non trivial condition
\begin{equation}\label{Bhorizon}
    \left(Q B\right)^2+8\left(Q B\right)+4=0, 
\end{equation}
whose solution is 
\begin{equation}\label{QBcondition1}
    QB=-4\pm2\sqrt{3}, 
\end{equation}
thus, the horizon coincidence requires
\begin{equation}
    QB<0.
\end{equation}}
{The same sign condition follows more generally from Eq. \eqref{photoregion}.
Solving it for $QB$ gives
\begin{equation}\label{QBcondition2}
QB = \frac{ 4A_1-r^2(3r^2-5Mr+2Q^2)B^2 } {8r\sqrt{A_0}}. 
\end{equation}
In the equatorial domain for $SLS$, $r_+<r\leq 2M$, one has 
\begin{equation}
A_0>0, \qquad A_1<0, \qquad 3r^2-5Mr+2Q^2>0. 
\end{equation} 
Therefore, the numerator in Eq. \eqref{QBcondition2} is negative, whereas the denominator is positive. Hence, to locate $r_{\rm ph}$ throughout this radial domain,  Eq.\ \eqref{photoregion} requires $QB<0$.}

{A second limiting configuration is obtained when the photon region reaches the largest equatorial value of the $SLS$, namely
\begin{equation}
r_{\rm ph}=r_{SLS}=2M.
\end{equation}
Thus, for the sign choice $B>0$ and $Q<0$, the two limiting admissible configurations are\footnote{An analogous configuration can occur with $B<0$ and $Q>0$.}
\begin{equation}
    r_{\rm ph}=r_+:\quad Q=-M, \quad B=\frac{2\left(2-\sqrt{3}\right)}{M},
\end{equation}
and
\begin{equation}
    r_{\rm ph}=r_{SLS}=2M:\quad Q = - 2^{-1/4}M, \quad B=\frac{\sqrt{2}-1}{M}.
\end{equation}
These two endpoints delimit the parameter interval for which the photon region lies between the horizon and the largest equatorial $SLS$, namely
\begin{equation}
    -M \leq Q\leq -2^{-1/4}M, \quad \frac{2\left(2-\sqrt{3}\right)}{M}\geq B \geq \frac{\sqrt{2}-1}{M}.
\end{equation}
Within this interval, the kinematical escape requirement can be consistently implemented as the lower bound in Eq.\ \eqref{radialwindow}.}

\begin{figure}[htbp!]
\includegraphics[scale=0.9]{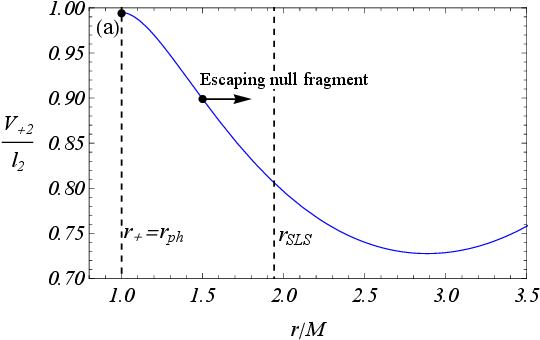}
\includegraphics[scale=0.9]{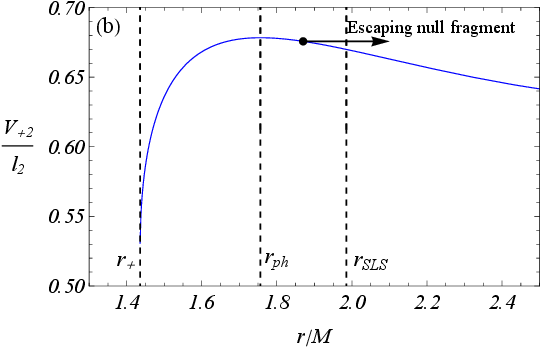}
\includegraphics[scale=0.9]{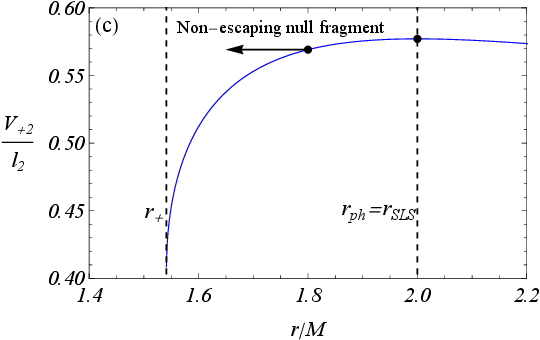}
\caption{\label{FIG16}Behavior of effective potential $V_{+2}/l_{2}$ for neutral particles in the equatorial plane for $M=1$. (a) Limiting configuration in which the photon region coincides with the event horizon, $r_{\rm ph}=r_+$. (b) Representative non-limiting configuration with $r_{\rm ph}<r_\ast<r_{SLS}$, showing a break-up point inside the allowed radial window for local extraction and escape. (c) Limiting configuration in which the photon region coincides with the $SLS$, $r_{\rm ph}=r_{SLS}$. The Figure illustrates how the relative position of $r_{\rm ph}$, $r_\ast$, and $r_{SLS}$ controls the escape of the positive energy fragment. The parameters and values of relevant radii are given in Table \ref{Tab4}.}
\end{figure}

{Therefore, the external magnetic field acts not only as a control parameter for the efficiency and for the size of the local extraction region, but also for the compatibility between local energy extraction and the subsequent escape of massless fragments. This compatibility is governed by the relative position of $r_{\rm ph}$, $r_\ast$, and $r_{SLS}$.}\\
{This behavior is illustrated in Fig.\ \ref{FIG16} through the effective potential $V_{+2}/l_2$. Figure\ \ref{FIG16}a shows the limiting case $r_{\rm ph}=r_+$, Fig.\ \ref{FIG16}b represents an intermediate configuration with $r_{\rm ph}<r_\ast<r_{SLS}$, and Fig.\ \ref{FIG16}c corresponds to the limiting case $r_{\rm ph}=r_{\rm SLS}$. The parameters and the relevant radii for each configuration are summarized in Table\ \ref{Tab4}.}

\begin{table}
\centering
\caption{Parameters and relevant radii for Fig.\ \ref{FIG16}.} \label{Tab4}
\begin{tabular}{c c c c c c}
\hline
             & $Q/M$      & $M B$  & $r_+/M$ & $r_{SLS}/M$ & $r_{\rm ph}/M$\\ \hline
a    & $-1$     & $2\left(2-\sqrt{3}\right)$& $1$ & $1.9404$ & $1$\\
b   & $- 0.9$       & $0.4751$ & $1.4358$& $1.9852$ & $1.7565$\\
c    & $- 2^{-1/4}$       & $\sqrt{2}-1$ & $1.5412$& $2$ & $2$\\
\hline
\end{tabular}
\end{table}

{For charged fragments, the boundary of the local extraction region is not necessarily identical to the $SLS$. While the $SLS$ is determined geometrically by $g_{tt}=0$, the boundary of the extraction region is fixed by the condition $\eta=0$ and therefore depends on the parameters of the charged particles and on the electromagnetic interaction. In this case, the previous kinematical argument can be generalized by replacing $r_{SLS}$ with the outer boundary $r_{eg}$ of the corresponding extraction region, defined by
\begin{equation}
\eta(r_{eg})=0.
\end{equation}
Thus, for a charged escaping fragment, the analogue of Eq.\ \eqref{radialwindow} becomes
\begin{equation}
r_{\rm esc}< r_\ast\leq r_{eg},
\end{equation}
where $r_{\rm esc}$ must be determined from the appropriate charged-particle effective potential. For the neutral massless case discussed above, this lower bound reduces to the photon-region threshold, $r_{\rm esc}=r_{\rm ph}$.}
\section{Conclusions}\label{sec4}
\vspace{-0.2cm}

In this work, the energy extraction process from a Reissner-Nordstr\"om BH immersed in an external uniform magnetic field is analyzed, as a natural extension of the study presented in \cite{Nucamendi2022}. We show that this system exhibits a much richer structure than the standard Kerr-like scenarios. The decay of uncharged and charged particles {at common turning points} of the radial motion is considered, leading to explicit expressions for the energies of the fragments in terms of the background geometry and the electromagnetic potential.  This formulation makes explicit the interplay between gravitational and electromagnetic contributions to the energy balance and provides  natural criteria to identify the regions where negative energy states, and therefore energy extraction, are possible. 

{It is important to stress that, because the magnetized Reissner-Nordstr\"om spacetime is not asymptotically flat, the efficiencies obtained here should be interpreted as local Killing energy efficiencies associated with the chosen stationary Killing field and electromagnetic gauge. Therefore, they should not be understood as ADM efficiencies measured by asymptotic observers at spatial infinity. Within this interpretation, the process describes a local splitting in which one fragment carries negative conserved Killing energy into the black hole, while the other fragment carries conserved Killing energy larger than that of the incident particle.}

We show that the magnetic field acts as a control parameter that governs both the existence of the ergosphere and the efficiency of the process. Analytical expressions for the critical magnetic fields that determine the onset and suppression of energy extraction are derived.

A central result of our analysis is that the extraction region is not determined exclusively by the background geometry. Instead, the magnetic field and charged particles acts as dynamical control parameters capable of enlarging, suppressing, or reconnecting disconnected extraction domains. In particular, the extraction condition becomes a coupled geometric and electromagnetic criterion, so that the existence of negative energy states is governed by the interplay between the metric contribution encoded in $g_{tt}$ and the electromagnetic interaction through $A_t$. As a consequence, energy extraction may occur even outside the geometrically defined ergoregion ($g_{tt}>0$), provided the electromagnetic contribution compensates the gravitational term. Furthermore, the magnetic field generates critical configuration where the structure of the extraction region changes qualitatively, leading to transitions between disconnected and connected efficiency windows in parameter space.

Our results also prove that the maximal efficiency of the process is not necessarily tied to the break-up point close to event horizon. Once the full radial structure is taken into account, the efficiency becomes a quantity controlled by the balance between the spacetime geometry and the electromagnetic interaction. Therefore, analyses restricted to the near-horizon regime capture only part of the physical structure of the extraction process.
 
{Finally, we distinguished the local energetic extraction criterion from the escape requirement. For the neutral null fragments considered here, this leads to the kinematical window
\begin{equation}
    r_{\rm ph}< r_\ast\leq r_{\rm SLS},
\end{equation}
which ensures that the break-up point lies inside the local extraction region and outside the photon region threshold. For charged fragments, the same reasoning requires replacing the outer boundary $r_{SLS}$ by the corresponding boundary of the extraction region, defined by $\eta=0$, while the lower escape threshold must be obtained from the appropriate charged-particle effective potential.}
 
This observation suggests that energy extraction processes are closely related to the dynamics of charged particles in magnetized spacetimes. Since the existence of negative energy states is tied to the effective radial structure, the framework developed here provides a natural starting point for connecting extraction mechanisms with particle trajectories and orbital dynamics of the accretion disk.

More generally, our results indicate that the interplay between electromagnetic fields and spacetime geometry plays a central role in determining where and how energy extraction can occur. In this sense, the present analysis offers a complementary perspective to other mechanisms such as the Blandford-Znajek process or magnetic reconnection scenarios, where electromagnetic fields also mediate the transfer of energy from compact objects.

More broadly, the framework developed here suggests a natural bridge between Penrose-like extraction mechanisms, charged particle dynamics, and electromagnetic energy transfer processes in strong gravity. In particular, the coexistence of rotational and electromagnetic extraction channels in Kerr–Newman spacetime may lead to qualitatively new regimes of energy extraction whose structure remains largely unexplored.

We expect that the present analysis provides a useful starting point for the study of these combined mechanisms in astrophysical relevant magnetized black hole environments.

\vspace{0.5cm}
\section{Acknowledgments}
\vspace{-0.2cm}
NB acknowledges partial support by SECIHTI Project CBF2023-2024-811. ICM acknowledges financial support of SNII-SECIHTI, Mexico, grant CVU No. 173252. AB acknowledges financial support by SECIHTI, Mexico, through the PhD Scholarship 814092.



\end{document}